\documentclass[10pt,journal,compsoc]{IEEEtran}

\usepackage{cite}
\usepackage{graphicx}
\usepackage{url}
\usepackage{amsmath}
\usepackage{amssymb}
 \usepackage[caption=false,font=footnotesize]{subfig}
\usepackage{epsfig}
\usepackage{booktabs}
\usepackage{multirow}

\usepackage{array}
\usepackage{color}
\usepackage{tabularx}
\usepackage{longtable}
\usepackage{tabu}
\usepackage{algorithmic}
\usepackage[ruled,linesnumbered]{algorithm2e}  
\usepackage{threeparttable}

\usepackage{tikz}
\usetikzlibrary{backgrounds,calc,chains,fit,matrix,positioning,shadows,shapes.misc,circuits.ee}
\usepackage{makecell}

\newcommand{\etal}{\textit{et al}.}
\newcommand{\ie}{\textit{i}.\textit{e}.}
\newcommand{\eg}{\textit{e}.\textit{g}.}

\begin{document}
\title{Active Fine-Tuning from gMAD Examples Improves Blind Image Quality Assessment}

\author{Zhihua~Wang,~\IEEEmembership{Student Member,~IEEE} and 
        Kede~Ma,~\IEEEmembership{Member,~IEEE}
\IEEEcompsocitemizethanks{\IEEEcompsocthanksitem The authors are with the Department
of Computer Science, City University of Hong Kong, Kowloon, Hong Kong (e-mail: zhihua.wang@my.cityu.edu.hk, kede.ma@cityu.edu.hk).}
}

\markboth{IEEE Transactions on Pattern Analysis and Machine Intelligence}%
{Shell \MakeLowercase{\textit{et al.}}: Bare Demo of IEEEtran.cls for Journals}

\IEEEtitleabstractindextext{
\begin{abstract}
The research in image quality assessment (IQA) has a long history, and significant progress has been made by leveraging recent advances in deep neural networks (DNNs). Despite high correlation numbers on existing IQA datasets, DNN-based models may be easily falsified in the group maximum differentiation (gMAD) competition. Here we show that gMAD examples can be used to improve blind IQA (BIQA) methods. Specifically, we first pre-train a DNN-based BIQA model using multiple noisy annotators, and fine-tune it on multiple synthetically distorted images, resulting in a ``top-performing'' baseline model. We then seek pairs of images by comparing the baseline model with a set of full-reference IQA methods in gMAD. The spotted gMAD examples are most likely to reveal the weaknesses of the baseline, and suggest potential ways for refinement. We query human quality annotations for the selected images in a well-controlled laboratory environment, and further fine-tune the baseline on the combination of human-rated images from gMAD and existing databases. This process may be iterated, enabling active fine-tuning from gMAD examples for BIQA. We demonstrate the feasibility of our active learning scheme on a large-scale unlabeled image set, and show that the fine-tuned quality model achieves improved generalizability in gMAD, without destroying performance on previously seen databases. 

\end{abstract}

}

\maketitle

\IEEEdisplaynontitleabstractindextext
\IEEEpeerreviewmaketitle

\IEEEraisesectionheading{\section{Introduction}}
\IEEEPARstart{A}{s} a fundamental problem in computational vision, objective image quality assessment (IQA) involves matching how humans perceive image distortions~\cite{wang2006modern}, and has been studied since 1970’s~\cite{mannos1974effects}. High quality prediction performance can be achieved by comparing a test image to its original counterpart, a setting known as full-reference IQA~\cite{daly1992visible}. Humans are able to perform quality evaluation without any reference at amazing speed and efficiency, and therefore it is reasonable to build computational models to accomplish a similar goal~\cite{wang2011}. The resulting blind IQA (BIQA) methods are applicable to a variety of image processing and computer vision tasks~\cite{kong2018no,ma2017learning}, where reference images may not exist. Moreover, the problem of BIQA itself provides an important test bed for our understanding of natural photographic images.

Early attempts to BIQA were distortion specific~\cite{marziliano2004perceptual,wang2002no}, {\color{black} which are essentially distortion visibility and severity measures}. For example, if JPEG compression is assumed, it is straightforward to make measurements to detect $8\times 8$ blocking artifacts. Later, general purpose solutions were developed based on models of natural scene statistics (NSS)~\cite{moorthy2011blind,mittal2012no,mittal2013making}. The underlying assumption is that sensory neurons are highly adapted  to the statistical properties of the
natural environment through both evolutionary and developmental processes~\cite{simoncelli2001natural}. It follows that a measure of  the destruction of ``naturalness'' can provide a good approximation to perceived image quality. NSS-based BIQA models often transform raw images to more compact and sparser representations~\cite{ahmed1974discrete,mallat1999wavelet,schwartz2001natural} so that the statistical regularities can be easily revealed and summarized using common probability models, \eg, generalized Gaussian distributions. This general methodology was widely practiced by BIQA models before 2015, some of which added a data-driven component, learning dictionaries~\cite{ye2012unsupervised} and quality-aware centroids~\cite{xue2013learning} directly from distorted patches.

In the past five years, 
 data-driven BIQA models~\cite{bosse2017deep,ma2017end} based on deep neural networks (DNNs) came to outperform knowledge-driven models based on NSS, in terms of correlation with human data on existing IQA databases~\cite{sheikh2006statistical,Ponomarenko201557}. These methods are built upon successive stages of convolution, nonlinear activation, and downsampling. Training such architectures with millions of parameters would require massive quality annotations in the form of mean opinion scores (MOSs), which are, however, largely lacking due to significant costs of performing large-scale subjective experiments. Several strategies have been proposed to compensate for the lack of human-rated data, including fine-tuning pre-trained networks~\cite{bianco2018use,zhang2018blind}, training on image patches~\cite{bosse2017deep}, exploiting degradation processes~\cite{ma2017end,liu2017rankiqa}, leveraging multiple noisy annotators~\cite{ma2019blind}, and combining IQA databases~\cite{zhang2019learning}.
 
\begin{figure*}[t]
\centering
    \subfloat[]{\includegraphics[width=0.24\textwidth]{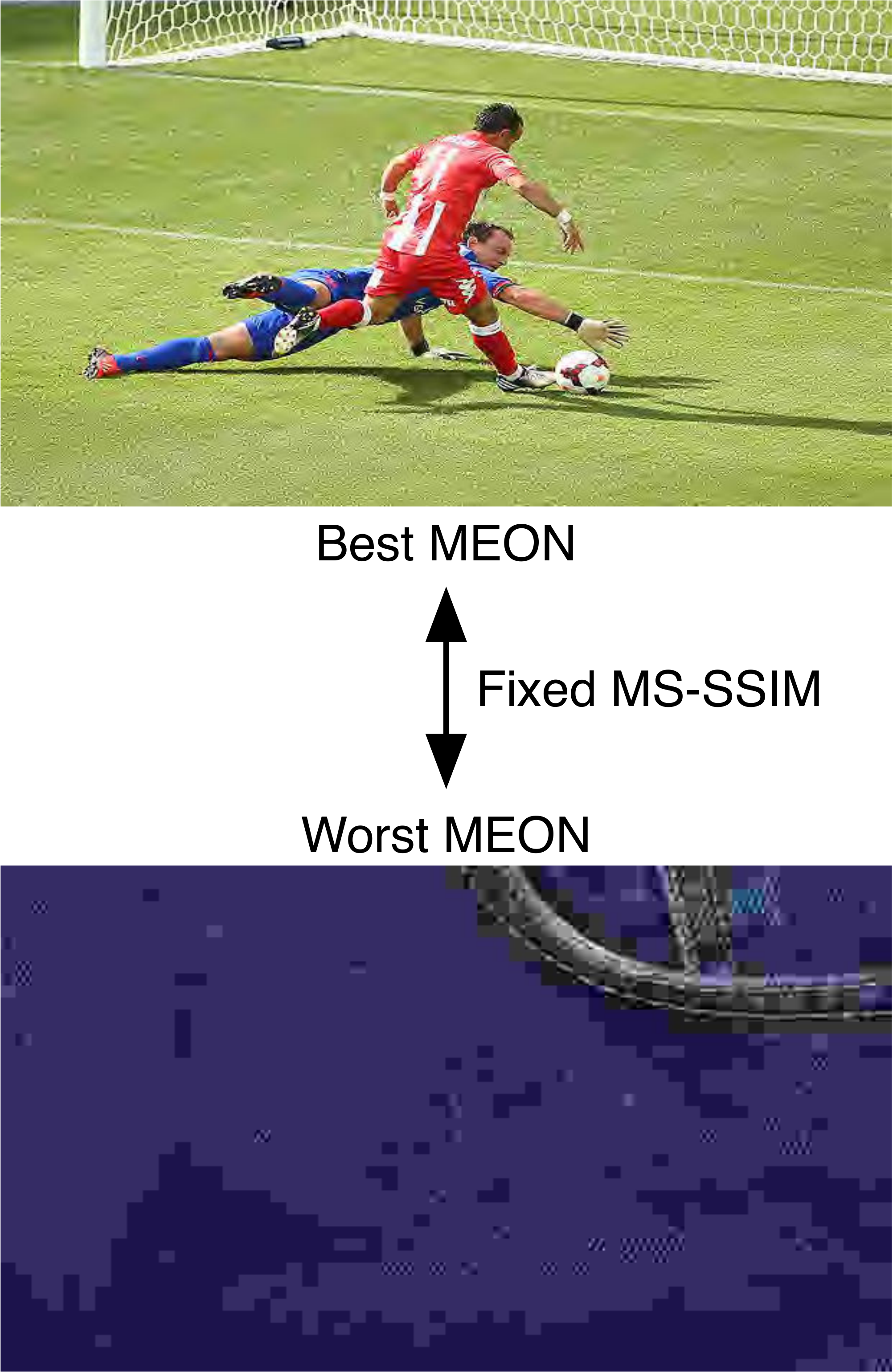}}\hskip.2em
    \subfloat[]{\includegraphics[width=0.24\textwidth]{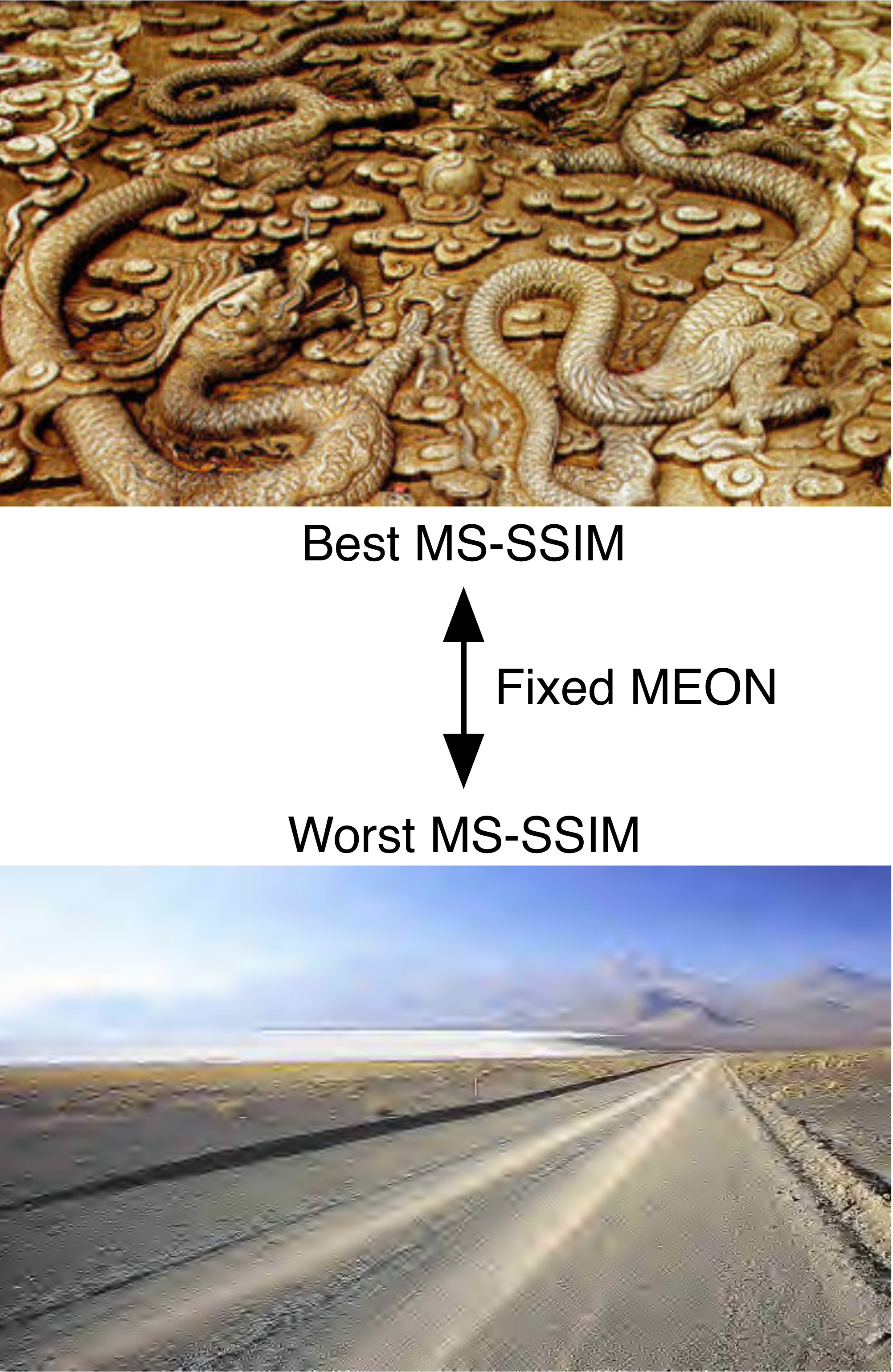}}\hskip.2em
    \subfloat[]{\includegraphics[width=0.24\textwidth]{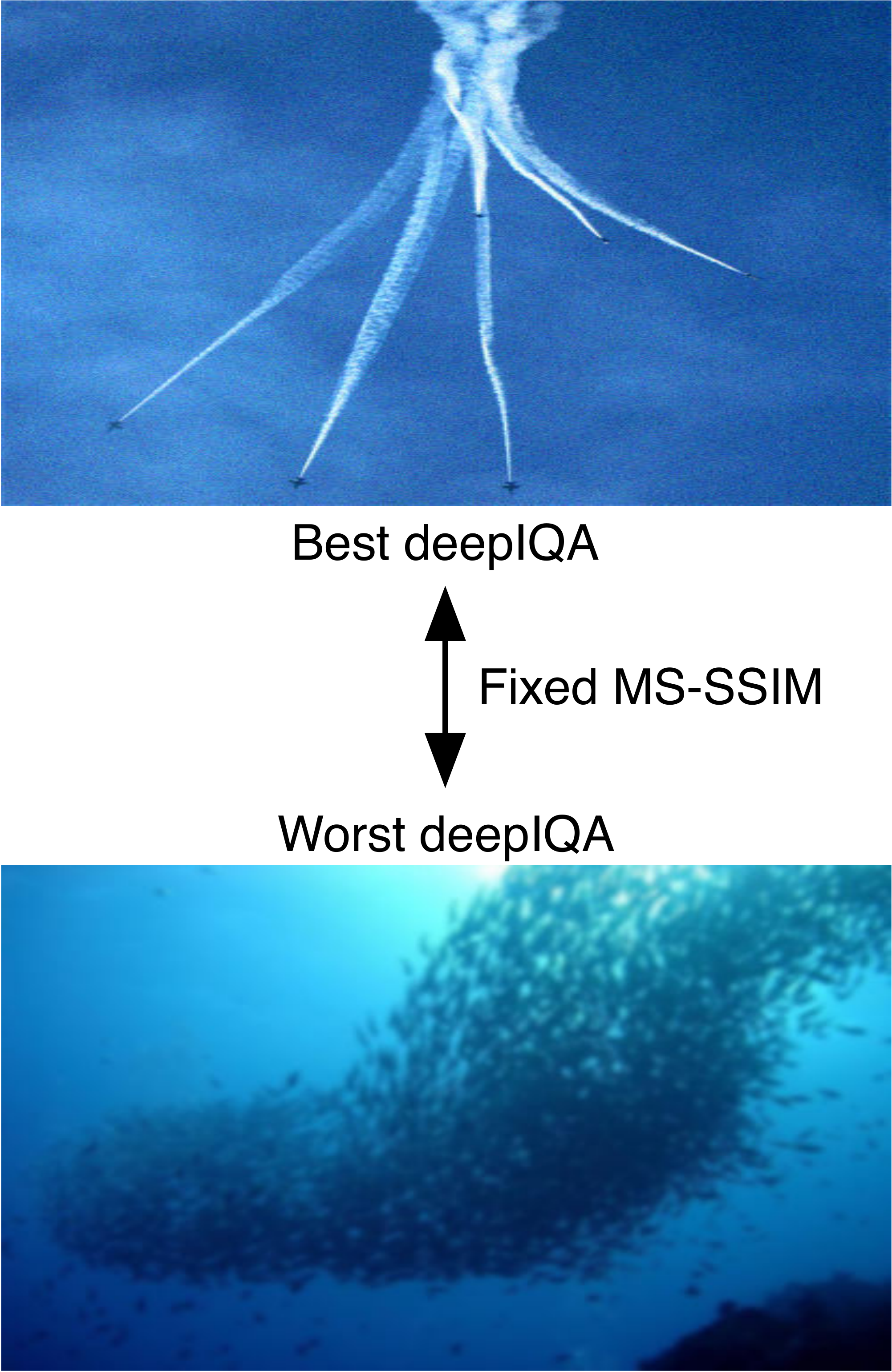}}\hskip.2em
    \subfloat[]{\includegraphics[width=0.24\textwidth]{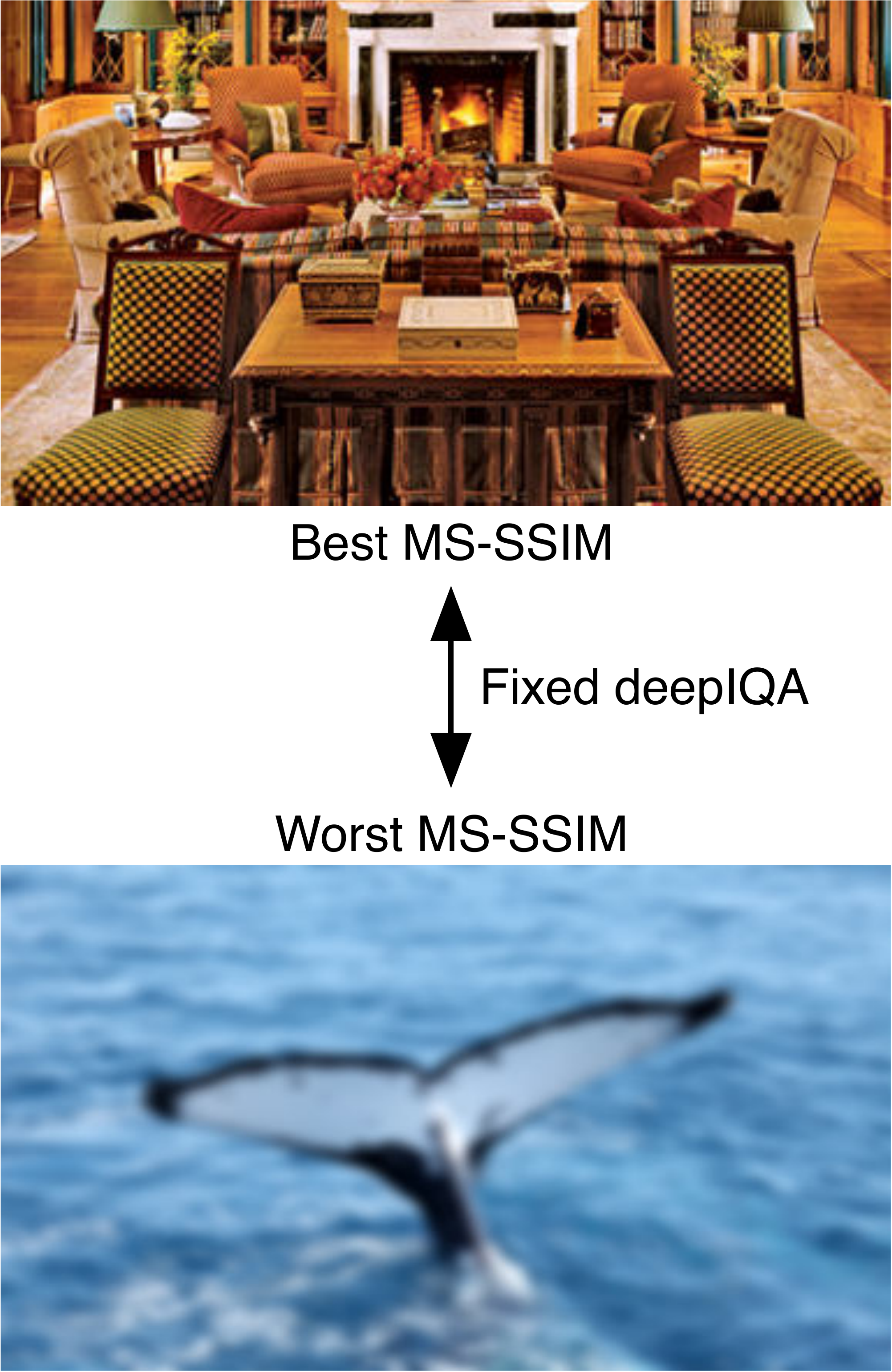}}
   
   \caption{Failures of two DNN-based BIQA models, MEON~\cite{ma2017end} and deepIQA~\cite{bosse2017deep}, when competing with a full-reference IQA method, MS-SSIM~\cite{wang2003multiscale}, in the gMAD competition on the Waterloo Exploration Database~\cite{ma2016waterloo}.  (a) Best/worst-quality images according to MEON, with near-identical quality reported by MS-SSIM. (b) Best/worst-quality images according to MS-SSIM with near-identical quality
reported by MEON. (c) Best/worst-quality images according to deepIQA with near-identical quality
reported by MS-SSIM. (d) Best/worst-quality images according to MS-SSIM with near-identical quality
reported by deepIQA. Visual inspection of the image pairs (a) and (b) indicates that MEON does not handle ringing artifacts well, which result from JPEG2000 compression. This suggests that exposing MEON to more diverse JPEG2000-compressed images during training may be a potential way of improving its robustness. Similarly, it is quite clear, from the images pairs (c) and (d), that deepIQA makes inaccurate quality predictions for  Gaussian-blurred images possibly due to its patch-based training strategy.}
\label{fig:intro_gmad_pairs}
\end{figure*}

Nevertheless, the impressive correlation numbers achieved by DNN-based BIQA models are questionable for two main reasons. First, model comparison has been performed using a small set of images, which are not sufficiently representative of the whole image population. Second, the same test images have been used to evaluate the models for many years.
This raises the risk of overfitting by extensive adaptation to excessively reused test sets. In fact, even for the best-performing BIQA models, dramatic failures can be found automatically via the group maximum differentiation (gMAD) competition~\cite{ma2018group}, a computational method of efficiently {\it falsifying} IQA models by  selecting
pairs of the most quality-discriminable images (see Fig.~\ref{fig:intro_gmad_pairs}).

In addition to testing the model generalizability using gMAD, here we shift our attention to leveraging gMAD examples to improve the BIQA performance. Focusing on predicting {\it relative} quality differences, 
we first pre-train a DNN-based BIQA model by learning from multiple noisy annotators~\cite{ma2019blind}, and then fine-tune it on four IQA databases simultaneously~\cite{zhang2019learning}. This gives rise to a top-performing baseline model that performs favorably against previous BIQA methods in assessing perceptual quality of synthetically distorted images. After that, the baseline is compared against nine {\color{black} stronger} full-reference IQA methods in  gMAD, attempting to seek its counterexamples for subjective testing. We further let the model adapt to the selected gMAD examples without forgetting previously trained databases by fine-tuning on images from both sources. Finally, we iterate the entire process of gMAD example mining, subjective testing, and fine-tuning several rounds, enabling active learning from gMAD examples for BIQA. \textcolor{black}{In summary, our contributions include:
\begin{itemize}
    \item A computational method to  efficiently expose and harness the failures of top-performing BIQA models for improved generalizability.
    \item A large-scale real experiment to demonstrate the feasibility of our active learning scheme. For example, the fine-tuned BIQA method eventually surpasses the nine full-reference IQA models in  gMAD.
\end{itemize}}

\section{Related Work}
In this section, we review previous work that is closely related to ours, including  DNN-based BIQA methods, gMAD competition, and machine learning from hard examples.

\subsection{DNNs for BIQA}
The main challenge to train DNNs for BIQA is that the small number of human-rated images may not be sufficient to constrain the large number of model parameters, typically in the order of millions. Directly fine-tuning pre-trained DNNs on image classification for BIQA seems a straightforward approach~\cite{bianco2018use,zhang2019learning}. However, it is unclear whether such network architectures and functional units are optimal for the BIQA task. Another strategy is to pre-train DNNs using quality-relevant data that can be generated efficiently. For example, Kang~\etal~\cite{Kang2015s}, Liu~\etal~\cite{liu2017rankiqa}, and Zhang~\etal~\cite{zhang2018blind} exploited the distortion type (and level) information to learn perceptually meaningful initializations. Kim~\etal~\cite{kim2018deep} and Ma~\etal~\cite{ma2019blind} made use of quality predictions from full-reference IQA models as pseudo ground truths. Methods of this kind hold much promise in handling synthetic distortions, on which they have been trained. It remains a challenge to develop distortion-unaware BIQA methods with good generalizability to unseen distortion types. In this work, we choose to predict {\it relative} quality, and combine the methods in~\cite{ma2019blind} and~\cite{zhang2019learning} to create a top-performing BIQA model (see Table~\ref{tab:cr}), as the starting point of our active learning for BIQA from gMAD examples.

\subsection{gMAD Competition}
gMAD~\cite{ma2018group} is a discrete instantiation of the maximum
differentiation (MAD) competition~\cite{wang2008maximum}, a general methodology for accelerating the comparison of perceptual models. Specifically, given two IQA models, MAD first synthesizes a pair of images by solving the following constrained optimization problem:
\begin{align} 
({x}^{\star}, y^{\star})& = \mathop{\text{argmax}}_{x,y} f_1(x) - f_1(y)\nonumber\\
&\text{ s.t. } f_2(x) = f_2(y)=\textcolor{black}{\xi}, \; x,y\in \mathcal{I},
\label{eq:mad}
\end{align}
where $f_j$, for $j=1,2$, are two objective quality models with larger values indicating better predicted quality, and $\mathcal{I}$ denotes the set of all possible images. The feasible image pairs are confined in the $\textcolor{black}{\xi}$-level set of $f_2$. By varying $\textcolor{black}{\xi}$, we are able to compare $f_1$ and $f_2$ at different quality levels. In Problem~(\ref{eq:mad}),
$f_1$ serves as an ``attacker'', whose difference of the responses to the pair of images $(x^\star, y^\star)$  is maximized, while $f_2$ works as a ``defender'', whose responses to $(x^\star, y^\star)$ are indistinguishable. MAD repeats this optimization, but with
the roles of the two models reversed~\cite{wang2008maximum}. The resulting small set of synthesized images constitutes the strongest possible examples to falsify the competing models~\cite{wang2008maximum}. 

However, MAD requires a projected gradient descent solver to synthesize images, which is computationally expensive, and is not friendly to non-differentiable IQA models. Moreover, the MAD-synthesized images may be highly unnatural, offering less insight into the relative model performance in real-world applications. gMAD overcomes the above limitations by restricting the search space to a fixed set of images $\mathcal{S}$, \ie, a particular \textit{domain of interest}. Efficient discrete optimizers can be adopted to solve Problem~(\ref{eq:mad}) to global optima. Based on subjective data, gMAD  introduces two quantitative measures, {\it aggressiveness} and {\it resistance}, to summarize
the performance of a model at attacking and defending against other models, respectively~\cite{ma2018group}. A number of  researchers~\cite{ma2017dipiq,yan2018two,zhang2018blind} have adopted gMAD to test the generalizability of their proposed models.  However, little work has been dedicated to exploiting gMAD examples to improve the  generalizability of BIQA models.

\subsection{Machine Learning from Hard Examples}
There is a rich body of literature on learning from hard examples, and the definition of ``hardness'' depends on the formulation and the goal of the machine learning task at hand.  In the case of hard negative mining~\cite{felzenszwalb2009object} (also referred to as bootstrapping~\cite{Sung:1996:LES:929901}), training is prioritized for samples with \textit{high loss} at each iteration. In the case of continual learning~\cite{ring1994continual} (also often called lifelong learning), the model tries to transfer knowledge learned from previous tasks to new ones with resistance to catastrophic forgetting. The hard examples are mainly from new tasks that may cause performance degradation of previously seen data. In the case of active learning~\cite{settles2009active}, the hard examples are generally \textit{informative} samples that the model is least certain or expects most change. Active learners aim to train on as few labeled instances as possible to achieve high performance, thereby minimizing the cost of labeling~\cite{settles2009active}. Our training paradigm can be seen as a form of active fine-tuning, where we actively seek informative samples for visual inspection by means of the gMAD competition. The selected examples are most likely to be the strongest possible counterexamples, which may lead to the greatest change to the model. However, the goal here is different: instead of minimizing the effort of subjective testing in IQA~\cite{ye2014active}, we aim to improve the generalizability of the BIQA model by learning from the selected gMAD examples.

\section{Proposed Method}
\begin{figure}[t]
\centering
   \includegraphics[width=1\linewidth]{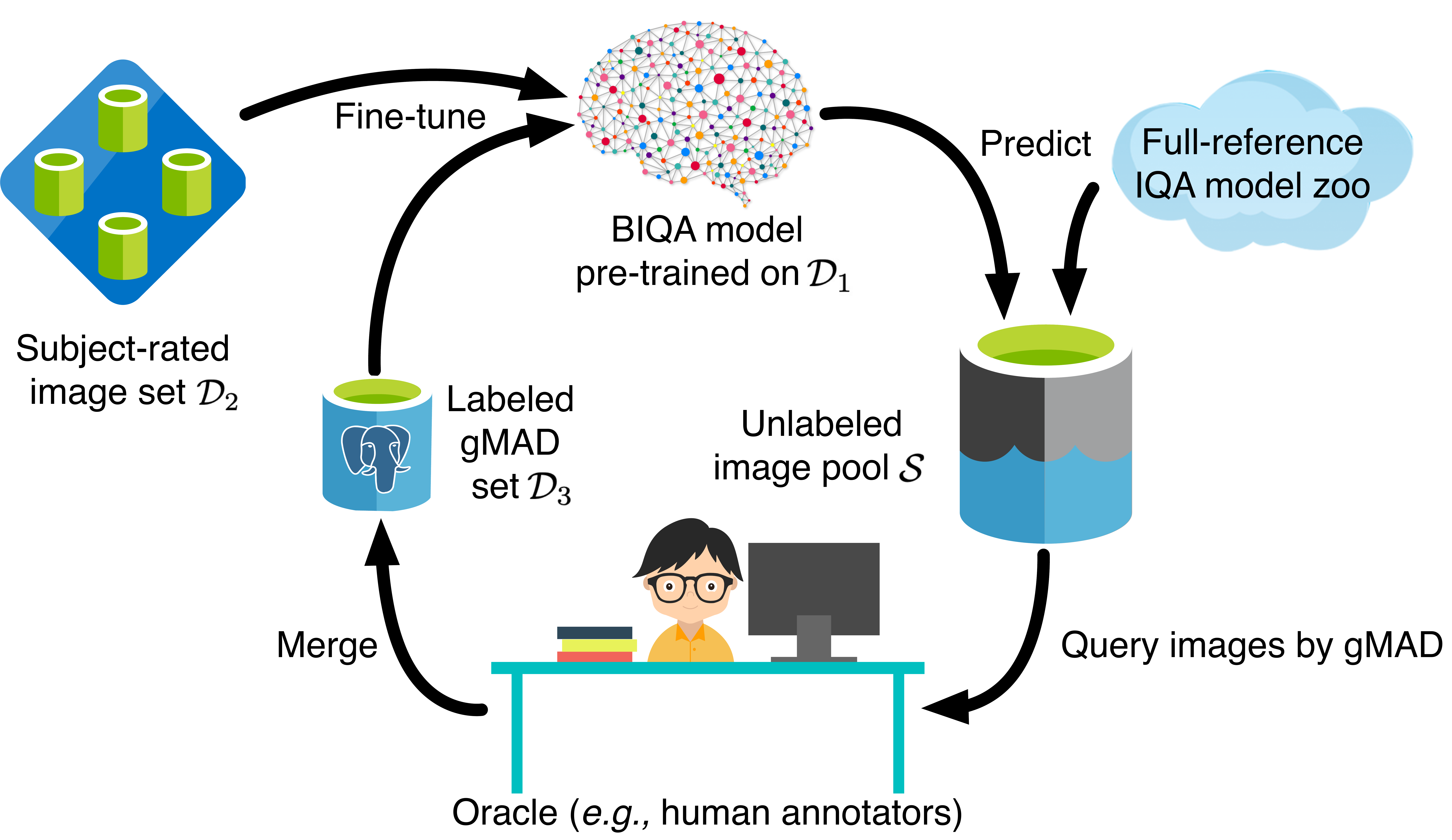}
\caption{The active fine-tuning cycle for improving BIQA models. We start with a differentiable parametric BIQA model, seek a small number of image pairs by letting it compete with a set of full-reference IQA methods in gMAD~\cite{ma2018group}, collect human opinions on the visual quality of the selected images, and fine-tune it from the combination of existing IQA databases and the newly annotated gMAD set.}
\label{fig:workflow}
\end{figure}

In this section, we describe the proposed method for BIQA, including baseline model construction followed by active  fine-tuning (see Fig.~\ref{fig:workflow}).  
\subsection{Constructing the Baseline Model}
 We build our baseline model in two steps: 1) pre-train a DNN on a large-scale database, with images annotated by a set of full-reference IQA methods~\cite{ma2019blind} and 2) fine-tune it on multiple IQA databases simultaneously~\cite{zhang2019learning}. The first step is used to supply perceptually meaningful initializations for subsequent fine-tuning.

 Given an image $x$, let $f({x})$ represent its true perceptual quality. We utilize $n$ IQA annotators $\{f_j\}_{j=1}^{n}$, which compute $n$ nonlinear and noisy quality estimates of $f(x)$, collectively denoted by $\{f_j({x})\}_{j=1}^{n}$. To cope with different model nonlinearities, an image pair $({x}, {y})$ is formed and associated with $n$ binary labels $\{q_j\}_{j=1}^n$, where $q_j = 1$ if   $f_j({x}) \ge f_j({y})$ and $q_j= 0$ otherwise. The training set is in the form of $\mathcal{D}_1=\{({x}^{(i)}, {y}^{(i)}), q^{(i)}_1, \ldots, q^{(i)}_n\}_{i=1}^m$, where $m$ is the number of training pairs. The reliability of each annotator is explicitly modeled by probabilities of correct answer and rejection rates: 
 \begin{align}
     \alpha_j= \Pr(q_j = 1 | q = 1) 
 \end{align}
 and
  \begin{align}
     \beta_j= \Pr(q_j = 0 | q = 0), 
 \end{align}
respectively, where $q=1$ if $f(x)\ge f(y)$ and $q=0$ otherwise.

 Our goal is to learn a differentiable function $f_w({x})$, parameterized by a vector $w$, which computes a quality value of $x$.  Assuming the Thurstone's Case V model~\cite{thurstone1927law}, the probability that $x$ is of higher quality than ${y}$ can be computed by
 \begin{align}
     p_w(x,y)&= \Pr(f({x})\ge f({y});{w})=\Phi\left(\frac{f_{{w}}({x}) - f_{{w}}({y})}{\sqrt{2}}\right),
         \label{eq:cdf}
\end{align}
where $\Phi(\cdot)$ is the standard Normal cumulative distribution function with standard deviation (std) fixed to one. The model parameters $w$ along with the uncertainty variables $\{\alpha, \beta\}$ are jointly estimated by maximum likelihood~\cite{ma2019blind}:

\begin{align} 
\{\hat{w}, \hat{\alpha},\hat{\beta}\} = \mathop{\text{argmax}}_{w,\alpha,\beta} \Pr(\mathcal{D}_1;w,\alpha,\beta),
\label{eq:ml19}
\end{align}
where 
\begin{align}
 \Pr(\mathcal{D}_1;w,\alpha,\beta) =& \prod_{i=1}^m\bigg(p_w(x^{(i)},y^{(i)})\prod_{j=1}^n\Pr(q_j^{(i)}|q=1)\nonumber\\
 +&(1-p_w(x^{(i)},y^{(i)}))\prod_{j=1}^n\Pr(q_j^{(i)}|q=0)\bigg).   
\end{align}
As shown in~\cite{ma2019blind}, the learned model is capable of handling distortion types that have been pre-specified in the training set $\mathcal{D}_1$, but does not generalize well to unseen distortions, especially those with substantially different visual appearances.

 To enhance model generalizability, we leverage the training technique proposed in~\cite{zhang2019learning}, and  fine-tune our BIQA model on multiple subject-rated IQA databases simultaneously. Given $n$ IQA databases, $m_j$ pairs of images $\{(x^{(i)}_j,y^{(i)}_j)\}_{i=1}^{m_j}$ are randomly sampled from the $j$-th database, and a total of $m=\sum_{j=1}^n m_j$ image pairs are constructed. For each pair $(x,y)$, a continuous quality annotation is computed, indicating the probability of $x$ having higher perceived quality than $y$:
\begin{align}
    p(x,y) = \Pr(f(x) \ge f(y)) = \Phi\left( \frac{\mu(x) -\mu(y)}{\sqrt{\sigma^2(x)+\sigma^2(y)}}\right),
    \label{eq:gt}
\end{align}
where the \textcolor{black}{same} Thurstone's model~\cite{thurstone1927law} is assumed, and $\mu(x)$ and $\sigma(x)$ are the MOS of $x$ and the corresponding std, respectively. The training  set is therefore in the form  of $\mathcal{D}_2=\{\{(x^{(i)}_j,y^{(i)}_j),p_j^{(i)}\}_{i=1}^{m_j}\}_{j=1}^n$. In~\cite{zhang2019learning}, the fidelity loss~\cite{tsai2007frank} is used to measure the similarity between two discrete probability distributions:
\begin{align}\label{eq:fidelity}
\ell(x, y, p;w)
=& 1 - \sqrt{p(x,y)p_w(x,y)}  \nonumber \\
&-\sqrt{(1-p(x,y))(1-p_w(x,y))}.
\end{align}
\textcolor{black}{The fidelity loss has a clear physical interpretation, and is used to  measure the difference between two states of a quantum~\cite{nielsen2002quantum}. 
In this paper, we will also use it to monitor the progress of our BIQA model and to help pick gMAD pairs for qualitative comparison.}

Finally, the model parameters $w$ are fine-tuned by minimizing the mean fidelity loss over the combined database $\mathcal{D}_2$
\begin{align}\label{eq:fidelityall}
\ell(\mathcal{D}_2;w)
= \frac{1}{\vert\mathcal{D}_2\vert}\sum_{i,j}\ell(x_j^{(i)}, y_j^{(i)}, p_j^{(i)};w),
\end{align}
where $\vert\mathcal{D}_2\vert$ denotes the cardinality of $\mathcal{D}_2$.


\subsection{Active Fine-Tuning from gMAD Examples}
After acquiring the baseline model $f_w$, we are able to actively fine-tune it using a small set of model-dependent images selected by  gMAD. We first build a large-scale unlabeled image set $\mathcal{S}$ as the playground for gMAD. As the size of the gMAD set  $\mathcal{U}$ subject to visual inspection is orthogonal to that of $\mathcal{S}$, we may make $\mathcal{S}$ arbitrarily large such that it spans a great variety of natural scenes, distortion types and levels. We assume a subjective assessment environment, where we can collect the MOS  
of $x\in\mathcal{S}$ and its corresponding std.  We also assume a set of full-reference IQA methods $\{f_j\}_{j=1}^n$, each of which takes a distorted image $x$ and its corresponding reference $x'$ as input, and computes an estimate of the true perceptual quality, $f_j(x)$, where we have omitted $x'$ in the parenthesis to keep the notation uncluttered. 
Fixing a quality level $\textcolor{black}{\xi}$, we first let our model and the $j$-th full reference IQA method be the defender and the attacker, respectively. The optimal pair of images in terms of discriminating $f_w$ and $f_j$ can be found by solving 
\begin{align} 
(x^r, y^r)& = \mathop{\text{argmax}}_{x,y} f_j(x) - f_j(y)\nonumber\\
&\text{ s.t. } f_w(x) = f_w(y)=\textcolor{black}{\xi}, \; x,y\in \mathcal{S},
\label{eq:mad2}
\end{align}
where the $j$-th full-reference method believes that $x^r$ has much better visual quality than $y^r$, while our model suggests that they are of approximately the same quality. The subjective result of $(x^r, y^r)$ roughly falls into three categories:
\begin{itemize}
\item \textbf{Case I}. $p(x^r, y^r) \approx 1$:  $x^r$ is indeed of better quality than $y^r$. In this case, $f_j$ makes a successful attack, identifying a counterexample of $f_w$. The selected pair of images contain constructive information about improving  $f_w$.  
\item \textbf{Case II}. $p(x^r, y^r) \approx 0.5$: $x^r$ and $y^r$ have very similar visual quality. In this case, $f_w$ survives the attack from $f_j$, which is in disagreement with human visual inspection. $(x^r, y^r)$ is informative in discriminating the two models, but may contribute less to performance improvement of $f_w$.
\item \textbf{Case III}. $p(x^r, y^r) \approx 0$: $y^r$ has better quality than $x^r$. In this case, $(x^r, y^r)$ is able to falsify both models, leading to a double-failure result. The selected pair is useful for the refinement of $f_w$.
\end{itemize}

\begin{algorithm}[t] \label{alg:af}
\SetAlgoLined
\KwIn{A pseudo-labeled image set $\mathcal{D}_1$, a combined subject-rated image set $\mathcal{D}_2$, an unlabeled image set $\mathcal{S}$, a randomly initialized BIQA model $f_w$ parameterized by a vector $w$, a group of full reference IQA models $\{f_j\}_{j=1}^n$, maximum round number $T$ for performing subjective experiments}
\KwOut{An actively fine-tuned BIQA model $f_{\hat{w}}$}
Pre-train $f_w$ on $\mathcal{D}_1$ by minimizing Eq.~(\ref{eq:ml19})\\
Fine-tune $f_w$ on $\mathcal{D}_2$ by minimizing Eq.~(\ref{eq:fidelityall})\\
Compute the responses of $\{f_j\}_{j=1}^n$ on $\mathcal{S}$ \\
$\mathcal{D}_3 \gets \emptyset$\\
\For{$t \gets 1$ \KwTo $T-1$}
{
    Compute the responses of $f_w$ on $\mathcal{S}$ \\
    Seek gMAD pairs of $f_w$ with the help of $\{f_j\}_{j=1}^n$ to form the unlabeled set $\mathcal{U}^{(t)}\subset\mathcal{S}$\\
    Request human opinions on the visual quality of each image in $\mathcal{U}^{(t)}$ to form $\mathcal{L}^{(t)}$\\
    Test the performance of $f_w$ on $\mathcal{L}^{(t)}$ \\
    $\mathcal{D}_3 \gets \mathcal{D}_3 \bigcup \mathcal{L}^{(t)}$ \\
    Augment $\mathcal{D}_3$ to form $\mathcal{D}'_3$\\
    Fine-tune $f_w$ on the combination of $\mathcal{D}_2$ and $\mathcal{D}'_3$ by minimizing Eq.~(\ref{eq:af})\\
    {\color{black} $\mathcal{S} \gets \mathcal{S} \backslash \mathcal{U}^{(t)}$}
}
Compute the responses of $f_w$ on $\mathcal{S}$ \\
Seek gMAD pairs of $f_w$ to form $\mathcal{U}^{(T)}$\\
Collect human opinions to form $\mathcal{L}^{(T)}$\\
Test the performance of $f_w$ on $\mathcal{L}^{(T)}$
\caption{Active fine-tuning from gMAD examples for BIQA}
\end{algorithm}

We then switch the roles of the two models, and seek an image pair $(x^a, y^a)$, to which the difference of the responses of $f_w$ is maximized in the {\color{black}$\xi$}-level set of $f_j$. That is, $f_w$ thinks $x^a$ is perceived much better than $y^a$, while $f_j$ considers they are indistinguishable in terms of image quality. Subjective testing on $(x^a, y^a)$ leads to three similar outcomes:
\begin{itemize}
\item \textbf{Case IV}. $p(x^a, y^a) \approx 1$: $x^a$ is of  clearly higher quality than $y^a$. In this case, $f_w$ successfully spots a counterexample of $f_j$. However,  $(x^a, y^a)$ may be less useful to further enhance $f_w$. 
\item \textbf{Case V}. $p(x^a, y^a) \approx 0.5$: $x^a$ and $y^a$ are of approximately the same quality. In this case, the attack by $f_w$ is not successful,  which exposes its own weakness when competing with $f_j$. $(x^a, y^a)$ can be used to improve $f_w$.
\item \textbf{Case VI}. $p(x^a, y^a) \approx 0$: $y^a$ has  clearly better quality than $x^a$. In this case, we reach a double-failure conclusion once again.  As the responses of $f_w$ to $(x^a, y^a)$ are opposite to human judgments, harnessing $(x^a, y^a)$ would impart the largest change to $f_w$.
\end{itemize}
For a relatively weak BIQA model, when competing with a group of full-reference IQA methods, the selected gMAD pairs are more likely to fall into Case I and Case V, which manifest themselves as strong gMAD counterexamples, and offer potential ways for enhancement. For a high-performance BIQA model (as is the case in our paper), we would expect to see some gMAD pairs belonging to Case II and Case IV (see Fig.~\ref{fig:probability distribution}).

In practice, we assume $l$ quality levels (\ie, {\color{black}$\xi$} can take on $l$ values), and for each quality level, we choose top-$k$ gMAD pairs with $k$ largest response differences computed by the objective in Problem~(\ref{eq:mad2}). We then reverse the roles of the two models, finding another top-$k$ gMAD pairs. After pairwise comparison with $n$ full-reference methods, we obtain an unlabeled gMAD set $\mathcal{U}$ that contains $2\times k\times l\times n$ pairs. We invite a number of subjects to rate each image $x\in\mathcal{U}$ in a well-controlled laboratory environment (see Section~\ref{sec:subjective} for details). The MOS $\mu(x)$ and the associated std $\sigma(x)$ can be computed accordingly. The ground truth annotation $p(x,y) \in [0,1]$ for a gMAD pair $(x,y)$ can also be derived using Eq.~(\ref{eq:gt}), leading to a labeled gMAD set $\mathcal{L}$ of the same size. After active fine-tuning on $\mathcal{L}$, we may iterate this process several rounds: leverage new knowledge acquired by $f_w$ to seek another set of gMAD examples, request human annotations for the selected images, and improve $f_w$ based on the labeled set. This gives us a progressively expanded gMAD set $\mathcal{D}_3 = \{\mathcal{L}^{(t)}\}_{t=1}^{T-1}$, which is in the form of  $\{(x^{(i)},y^{(i)}), p^{(i)}\}_{i=1}^m$, where $m = 2\times k\times l\times n\times (T-1)$ and $T$ is the maximum number of rounds. Note that we reserve $\mathcal{L}^{(T)}$ for testing purpose only.

\begin{table*}
\caption{Summary of IQA databases. MOS stands
for mean opinion score. DMOS is inversely proportional to MOS}
\centering
\begin{tabular}{l|cccccc}
\toprule
\multirow{2}{*}{Database} & \# of original & \# of distorted & \# of distortion   & \multirow{2}{*}{Score type} & \multirow{2}{*}{Score range}  &\multirow{2}{*}{Subjective testing methodology} \\
& images & images & types & &&\\
\hline
LIVE~\cite{sheikh2006statistical} & $29$ & $779$ & $5$  & DMOS & $[0,100]$ & Single-stimulus continuous scale\\
CSIQ~\cite{larson2010most} & $30$ & $866$ & $6$  & DMOS & $[0, 1]$ & Multi-stimulus absolute category\\
TID2013~\cite{Ponomarenko201557} & $25$ & $3,000$ & $24$ & MOS & $[0, 9]$ & Two-alternative forced choice \\
KADID-10k~\cite{lin2019kadid} & $81$ & $10,125$ & $25$ & MOS & $[1, 5]$ & Double-stimulus absolute category\\
Waterloo Exploration~\cite{ma2016waterloo} & $4,744$ & $94,800$ & $4$ & N.A. & N.A. & Need-based\\
\bottomrule
\end{tabular}
\label{tab:dc}
\end{table*}

We now describe the $t$-th round of the fine-tuning procedure using the combination of image pairs from $\mathcal{D}_2$ and $\mathcal{D}_3$, where $\mathcal{D}_3 = \{\mathcal{L}^{(t')}\}_{t'=1}^{t}$. The goal is to
harness gMAD examples without overfitting, and preserve performance on  previously trained IQA databases. In general, the size of $\mathcal{D}_3$ is much smaller compared to that of $\mathcal{D}_2$. We alleviate this data imbalance in two ways. First, instead of directly adapting to the selected gMAD pairs, we randomly pair up gMAD images, which results in an augmented training set $\mathcal{D}'_3$ containing $m\times (2m+1)$ pairs. Second, we weight the loss function according to the number of instances in the respective databases:
\begin{align}\label{eq:af}
\ell(\mathcal{D}_2,\mathcal{D}_3;w)
&= \frac{1}{\vert\mathcal{D}_2\vert}\sum_{i,j}\ell(x_j^{(i)}, y_j^{(i)}, p_j^{(i)};w)\nonumber\\ 
&+ \frac{1}{\vert\mathcal{D}_3'\vert}\sum_{i}\ell(x^{(i)}, y^{(i)}, p^{(i)};w).
\end{align}
Algorithm~\ref{alg:af} summarizes the entire procedure of the proposed method. 



\section{Experiments}
In this section, we demonstrate the feasibility of the proposed method in real settings. We first present in  detail the baseline BIQA model for synthetic distortions. We then describe the active fine-tuning cycle, including the construction of the large-scale unlabeled image set $\mathcal{S}$, the implementation of the gMAD competition, the environment of the subjective experiment, and the procedure of active fine-tuning. Last, we conduct both quantitative and qualitative analysis of the proposed method with  a number of interesting observations. 
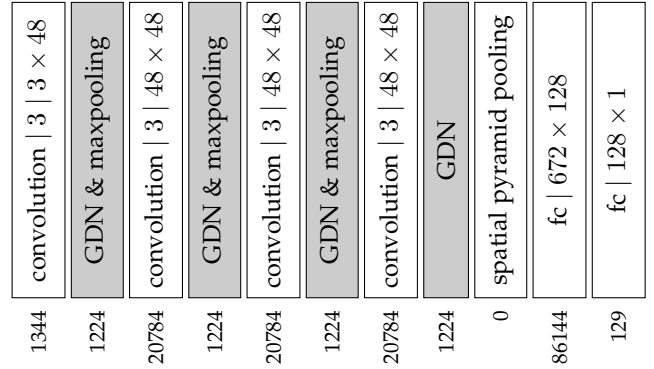
\begin{figure}[t]
\centering
\begin{tikzpicture}
\node [matrix, inner sep=0pt, column sep=2pt, row sep=1ex,
layer/.style={rectangle, draw, fill=white, rotate=90, minimum width=25ex, inner ysep=5pt},
numpars/.style={rotate=90, font=\footnotesize, left}] (analysis) {
\node[layer] {convolution $|\; 3 \;|\;3 \times 48$}; &
\node[layer, fill=black!20] {GDN \& maxpooling}; &
\node[layer] {convolution $|\; 3 \;|\;48\times 48$}; &
\node[layer, fill=black!20] {GDN \& maxpooling}; &
\node[layer] {convolution $|\; 3 \;|\;48\times 48$}; &
\node[layer, fill=black!20] {GDN \& maxpooling}; &
\node[layer] {convolution $|\; 3\;|\;48\times 48$}; &
\node[layer, fill=black!20] {GDN}; &
\node[layer] {spatial pyramid pooling}; &
\node[layer] {fc $|\;672\times 128$}; &
\node[layer] {fc $|\; 128\times 1$}; &\\

\node[numpars] {1344}; &
\node[numpars] {1224}; &
\node[numpars] {20784}; &
\node[numpars] {1224}; &
\node[numpars] {20784}; &
\node[numpars] {1224}; &
\node[numpars] {20784}; &
\node[numpars] {1224}; &
\node[numpars] {0}; &
\node[numpars] {86144}; &
\node[numpars] {129}; &
\\
};
\end{tikzpicture}
\caption{The network architecture of our BIQA model. The parameterization of convolution is denoted as ``filter support $|$ input channel $\times$ output channel.'' The number of parameters for each layer is given at the bottom, summing up to $154,865$.}
\label{fig:dnn_architecture}
\end{figure}

\subsection{Specification of the Baseline Model}
\label{section:basemodel}
\subsubsection{Network Architecture}Our BIQA model is adapted  from~\cite{ma2019blind} and is specified in Fig.~\ref{fig:dnn_architecture}. $f_w$ is a four-layer convolutional network. Each layer applies a bank of $3\times 3$ convolutional filters to its inputs. Following each convolution,
we employ generalized divisive normalization (GDN), in which all responses are divided by pooled responses of their rectified and exponentiated neighbors~\cite{Balle16a}. It implements a form of local gain control, which is useful in explaining  nonlinear behaviors
of cortical neurons~\cite{heeger1992normalization}. GDN is defined as
\begin{align} v_{i} = \frac {u_{i}}{\left ({\omega_{i} + \sum _{j}\gamma _{ij}u_{j}^{2}}\right )^{\frac {1}{2}}},
\end{align}
where $u$ and $v$ are the input to and the output of GDN, respectively, and $\{\omega,\gamma\}$ are the parameters to be determined. Apart from IQA~\cite{laparra2016perceptual,ma2017end}, GDN has also been successfully adopted in density modeling~\cite{Balle16a} and image compression~\cite{Balle17a}. The normalization responses are max-pooled by a factor of two along each spatial dimension. The  spatial statistics are summarized using spatial pyramid pooling~\cite{he2014spatial},
\textcolor{black}{ which hierarchically pools the features using local spatial bins, and generates a fixed-length representation regardless of input image resolution}. Last, the quality value is computed by two fully connected layers with a rectified linear unit (ReLU) in between.

\subsubsection{Construction of $\mathcal{D}_1$}We build the pseudo-labeled image set $\mathcal{D}_1$ based on the reference images from the Waterloo Exploration Database~\cite{ma2016waterloo}. We simulate $18$ common distortions\footnote{These include additive white Gaussian noise, multiplicative noise, pink noise, salt and pepper noise, Gaussian blur, JPEG compression, JPEG2000 compression, Gaussian denoising, color quantization, dithering, neighboring  patch substitution, flat patch substitution, contrast change, saturation decrease,  chromatic aberration, over-exposure, under-exposure, and ghosting.}, each at five levels. We assemble four types of image pairs~\cite{ma2019blind}: same reference
image and distortion type, with different distortion levels; same
reference image, but different distortion types and levels; two different reference images, distortion types and levels;  two different reference images, with one undistorted. We generate a total of $600,000$ training pairs, whose  labels are supplied by six full-reference IQA models.

\begin{table}[t]
\caption{Correlation (SRCC and PLCC) between model predictions and MOSs on $\mathcal{T}$. Top section lists two representative full-reference models. Second section contains four knowledge-driven and three data-driven DNN-based BIQA models. The results on the databases used to train the respective models are not shown. The top two correlations obtained by BIQA models are highlighted in boldface}
\centering
\begin{tabular}{l|cccc}
\toprule
SRCC & LIVE  & CSIQ & TID2013 & KADID-10k\\
\hline
SSIM~\cite{wang2004image} & $0.951$ & $0.871$ & $0.719$ & $0.747$\\
PieAPP~\cite{prashnani2018pieapp} & $0.919$ & $0.891$ & $0.885$ & $0.886$\\
\hline
BRISQUE~\cite{mittal2012no} & $-$ & $0.558$ & $0.407$ & $0.335$\\
NIQE~\cite{mittal2013making} & $\bf 0.922$ & $0.618$ & $0.315$ & $0.404$ \\
HOSA~\cite{xu2016blind} & $-$ & $0.602$ & $0.469$ & $0.353$\\
dipIQ~\cite{ma2017dipiq} & $\bf 0.944$ & $0.561$ & $0.412$ & $0.293$ \\
MEON~\cite{ma2017end}& $-$ & $0.741$ & $0.379$ & $0.214$\\
NIMA~\cite{talebi2018nima} & $0.506$ & $0.521$ & $0.301$ & $0.233$ \\
deepIQA~\cite{bosse2017deep} & $ 0.807$ & $0.752$ & $-$ & $0.595$\\
\hline
Baseline ($\mathcal{D}_1$) & $0.910$ & $\bf 0.870$ & $\bf 0.675$ & $\bf 0.621$ \\
Baseline ($\mathcal{D}_2$)   & $0.896$ & $\bf 0.859$ & $\bf 0.822$ & $\bf 0.861$ \\
\midrule
\midrule
PLCC& LIVE  & CSIQ & TID2013 & KADID-10k\\
\hline
SSIM & $0.940$ & $0.861$ & $0.784$ & $0.738$\\
PieAPP & $0.902$ & $0.880$ & $0.876$ & $0.887$\\
\hline
BRISQUE & $-$ & $0.677$ & $0.544$ & $0.394$\\
NIQE & $\bf 0.919$ & $0.742$ & $0.427$ & $0.460$ \\
HOSA & $-$ & $0.760$ & $0.590$ & $0.436$\\
dipIQ & $\bf 0.945$ & $0.758$ & $0.454$ & $0.400$ \\
MEON  & $-$ & $0.786$ & $0.486$ & $0.403$\\
NIMA & $0.511$ & $0.601$ & $0.476$ & $0.348$ \\
deepIQA & $0.839$ & $0.814$ & $-$ & $0.612$ \\
\hline
Baseline ($\mathcal{D}_1$) & $0.910$ & $\bf 0.902$ & $\bf 0.711$ & $\bf 0.628$ \\
Baseline ($\mathcal{D}_2$) & $0.915$ & $\bf 0.897$ & $\bf 0.837$ & $\bf 0.866$ \\
\bottomrule
\end{tabular}
\label{tab:cr}
\end{table}

\begin{figure*}[t]
\centering
\begin{minipage}[t]{1.0\textwidth}
    \centering
    \subfloat[]{\includegraphics[width=0.23\textwidth]{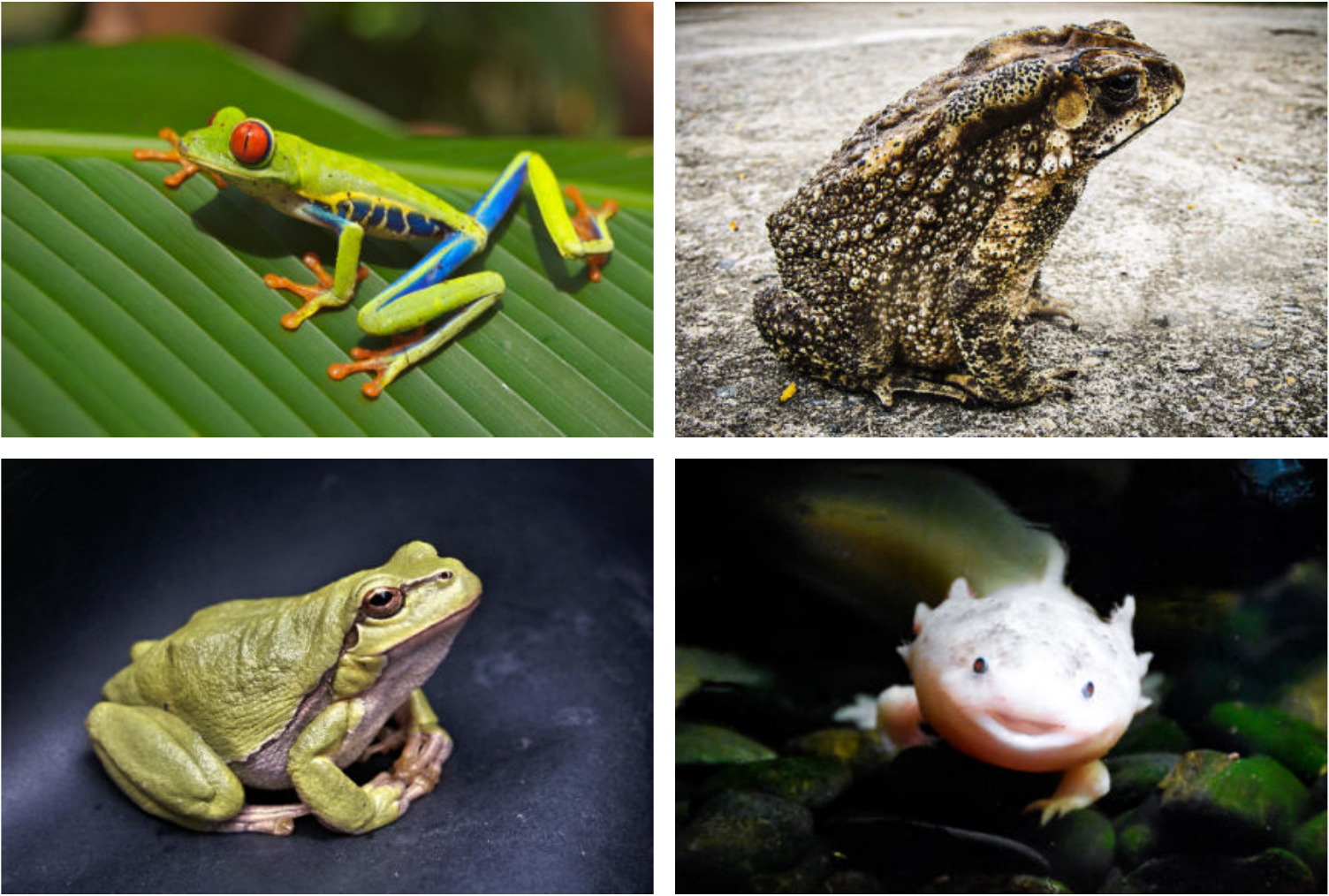}}\hskip.8em
    \subfloat[]{\includegraphics[width=0.23\textwidth]{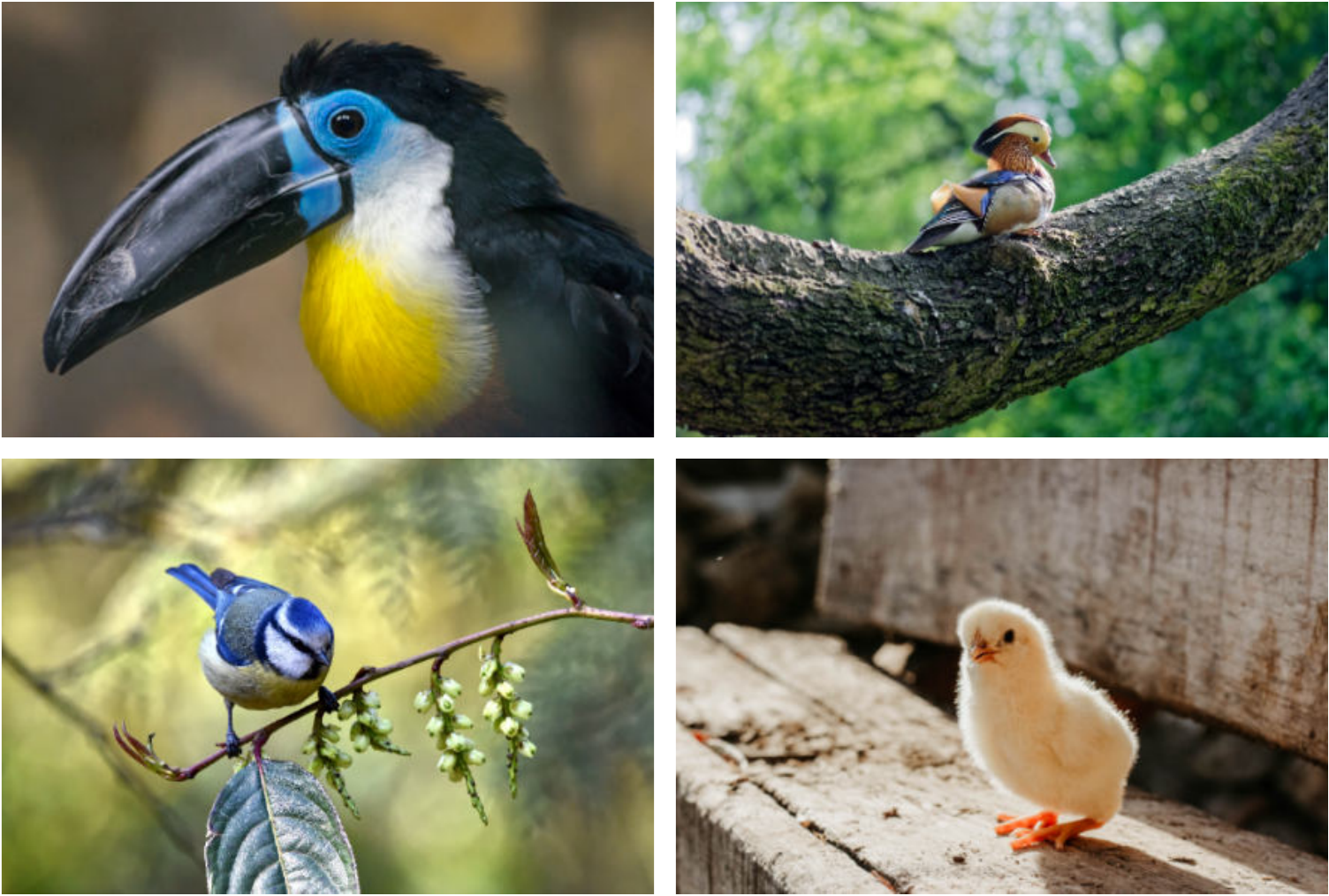}}\hskip.8em
    \subfloat[]{\includegraphics[width=0.23\textwidth]{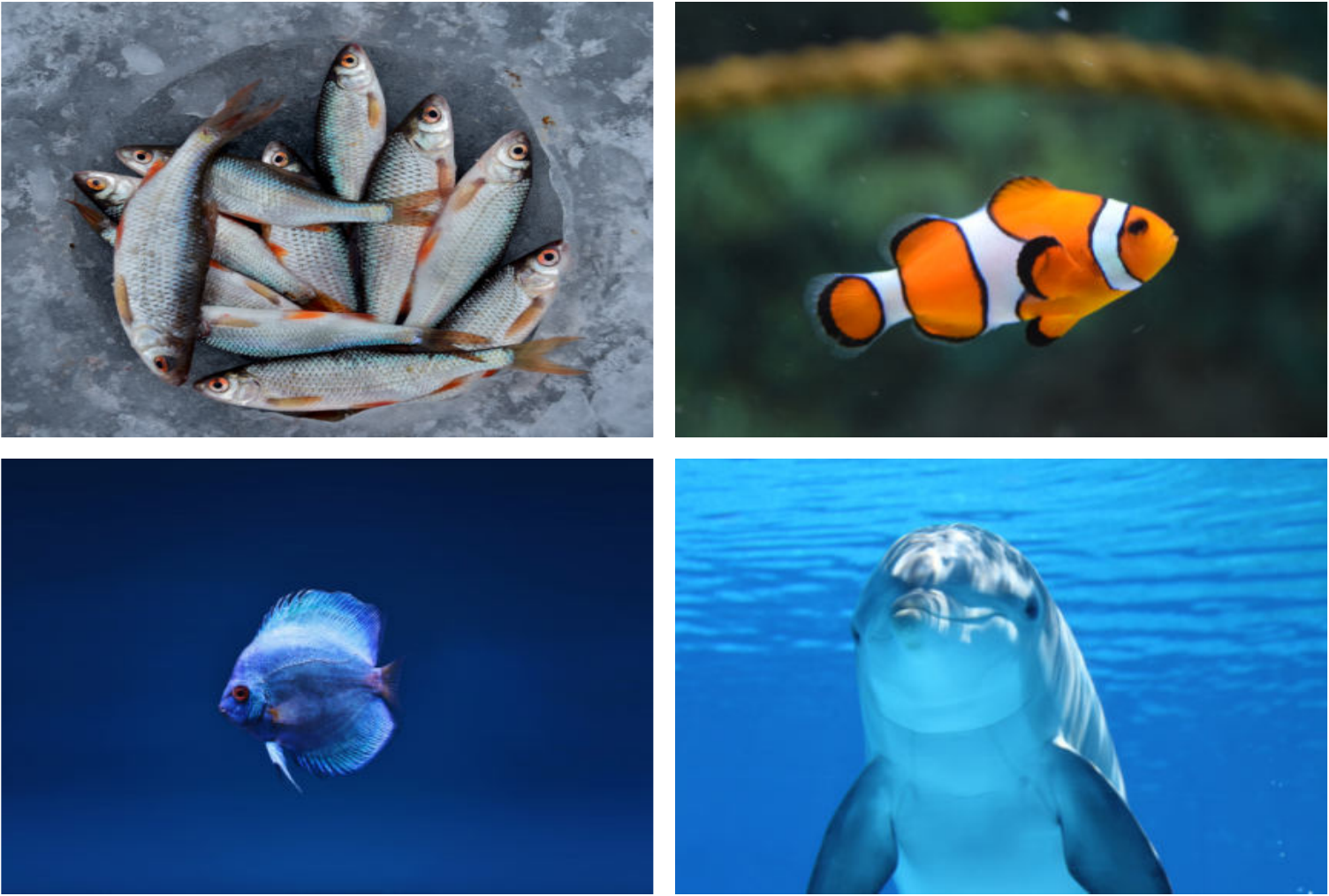}}\hskip.8em
    \subfloat[]{\includegraphics[width=0.23\textwidth]{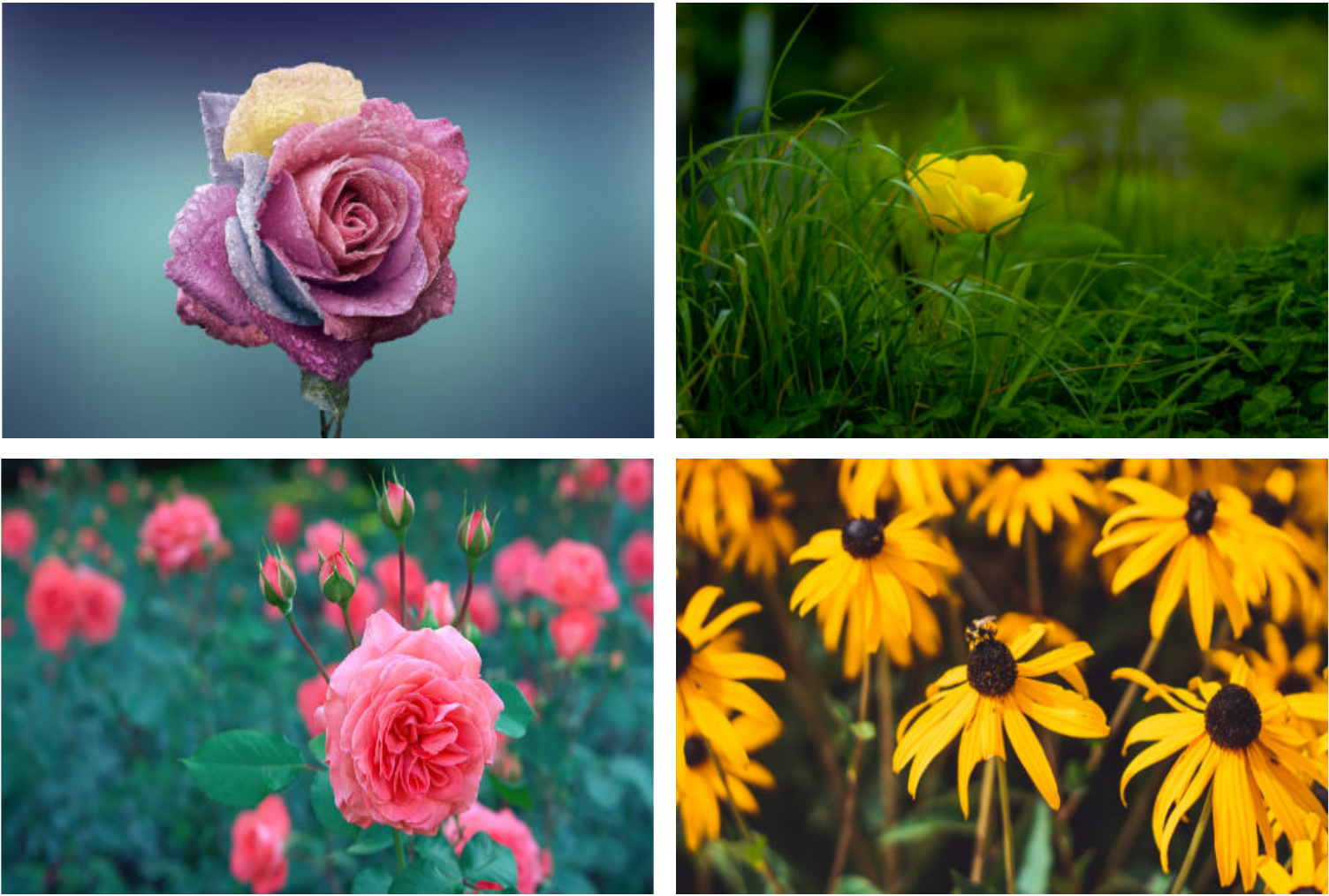}}\hskip.8em
\end{minipage}
\begin{minipage}[t]{1.0\textwidth}
    \centering
    \vspace{.2em}
    \subfloat[]{\includegraphics[width=0.23\textwidth]{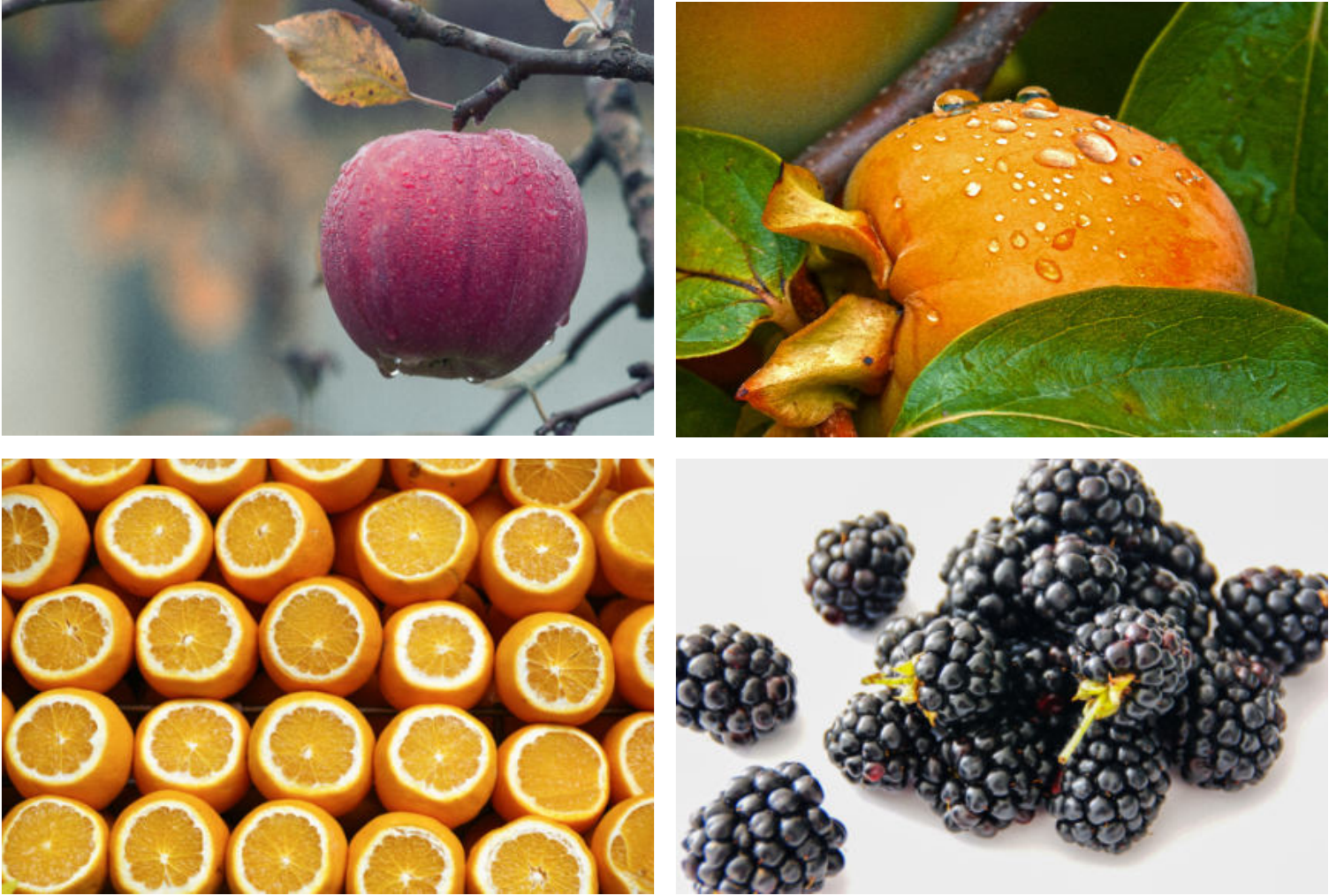}}\hskip.8em 
    \subfloat[]{\includegraphics[width=0.23\textwidth]{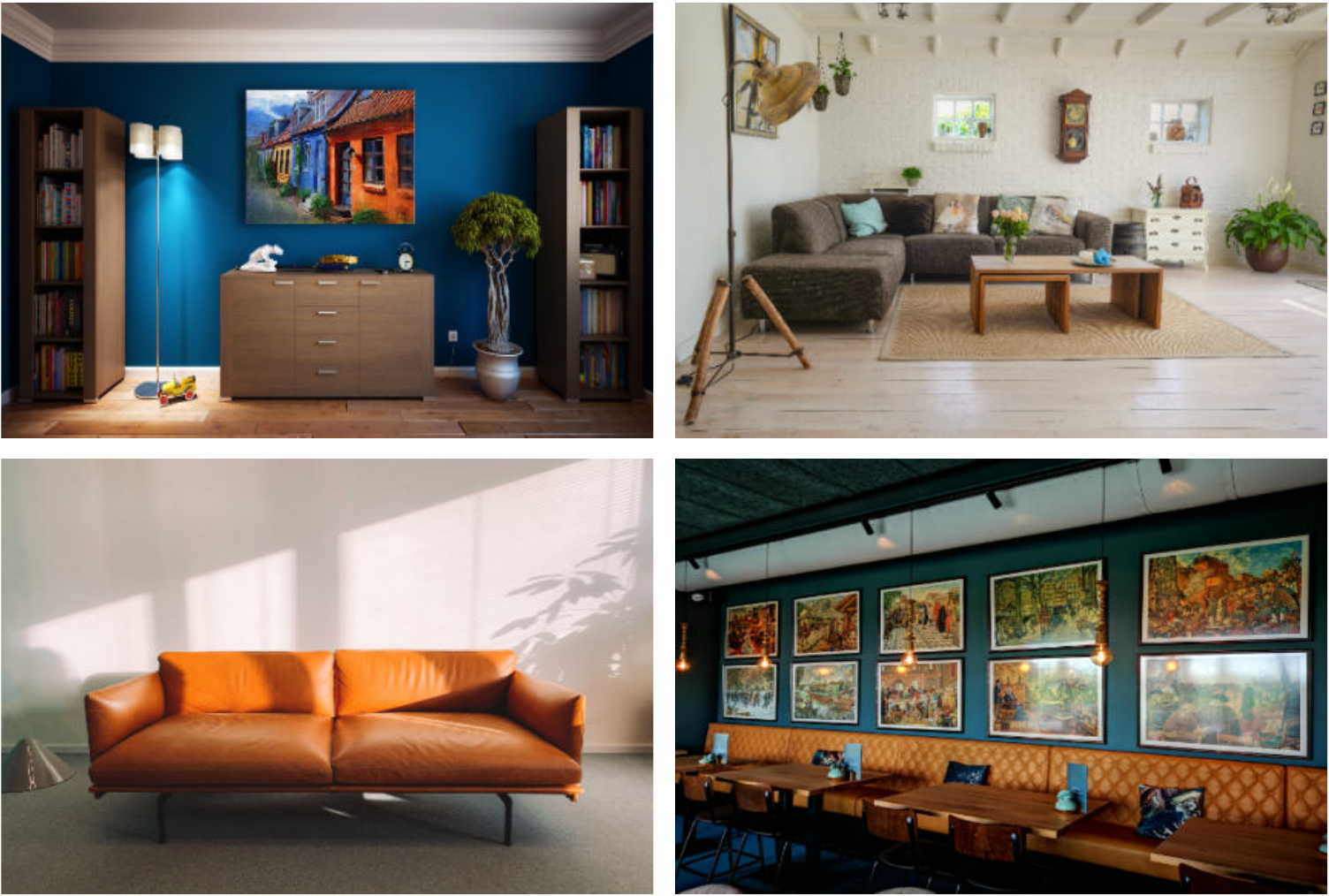}}\hskip.8em 
    \subfloat[]{\includegraphics[width=0.23\textwidth]{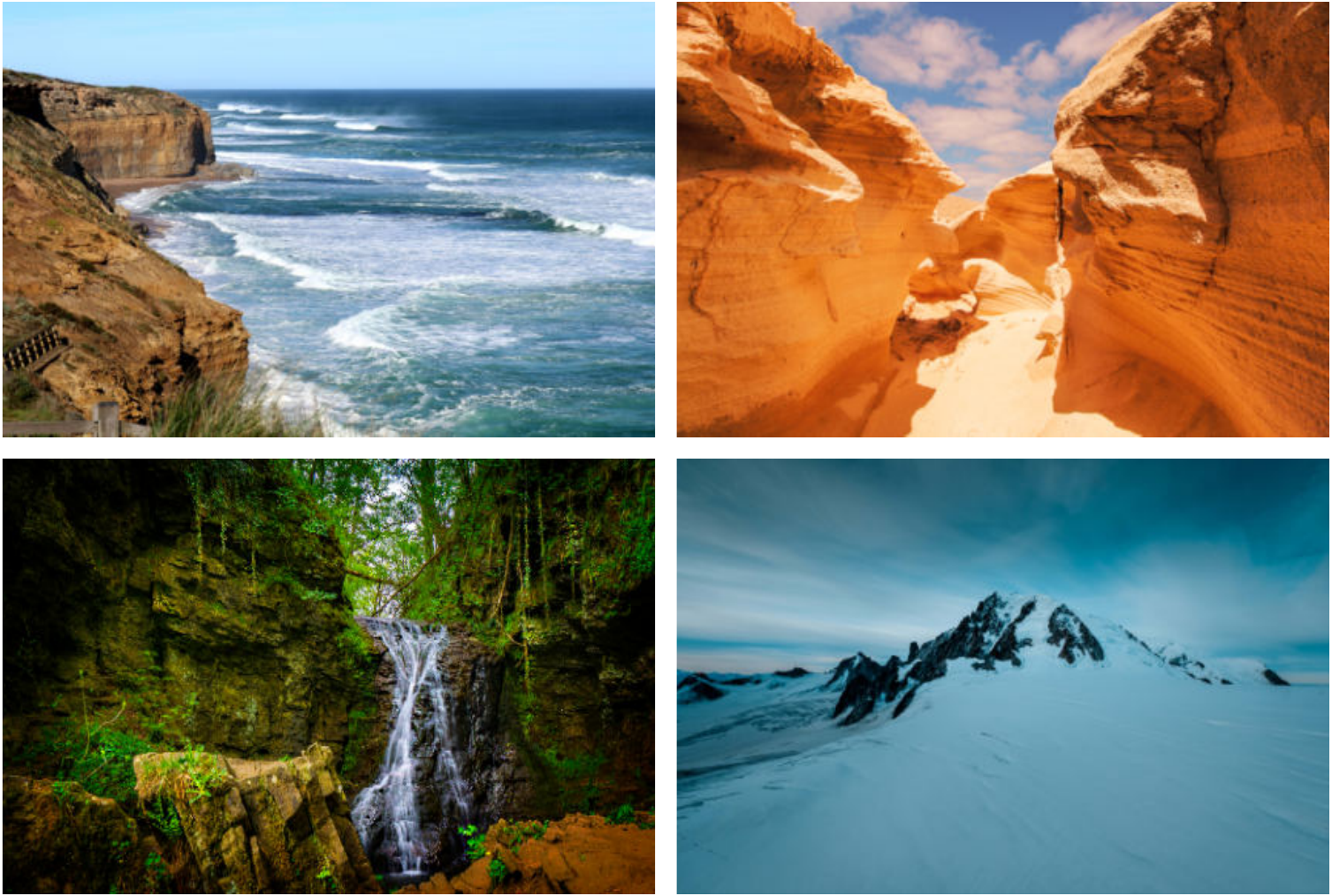}}\hskip.8em
    \subfloat[]{\includegraphics[width=0.23\textwidth]{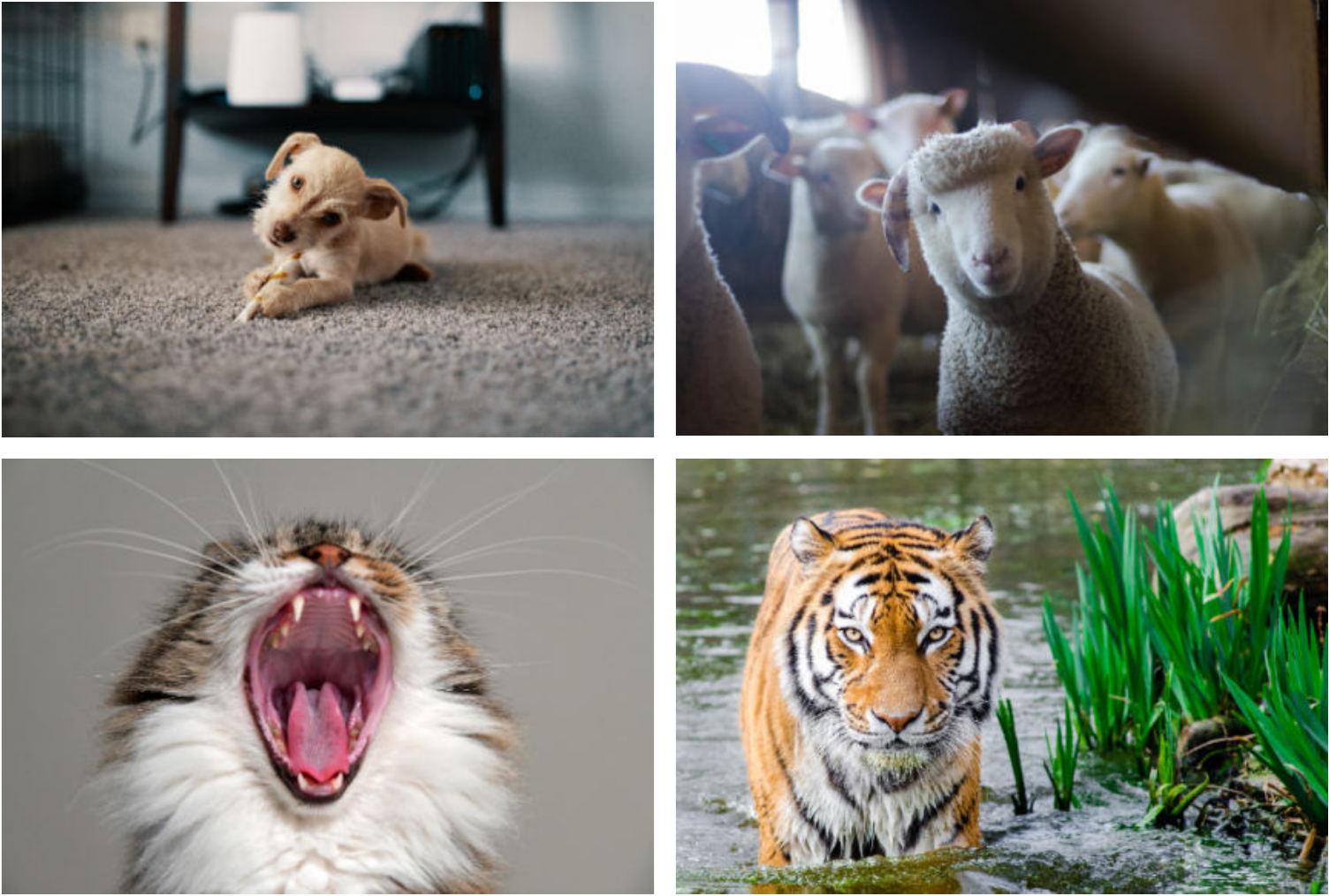}}\hskip.8em 
\end{minipage}
\begin{minipage}[t]{1.0\textwidth}
    \centering
    \vspace{.2em}
    \subfloat[]{\includegraphics[width=0.23\textwidth]{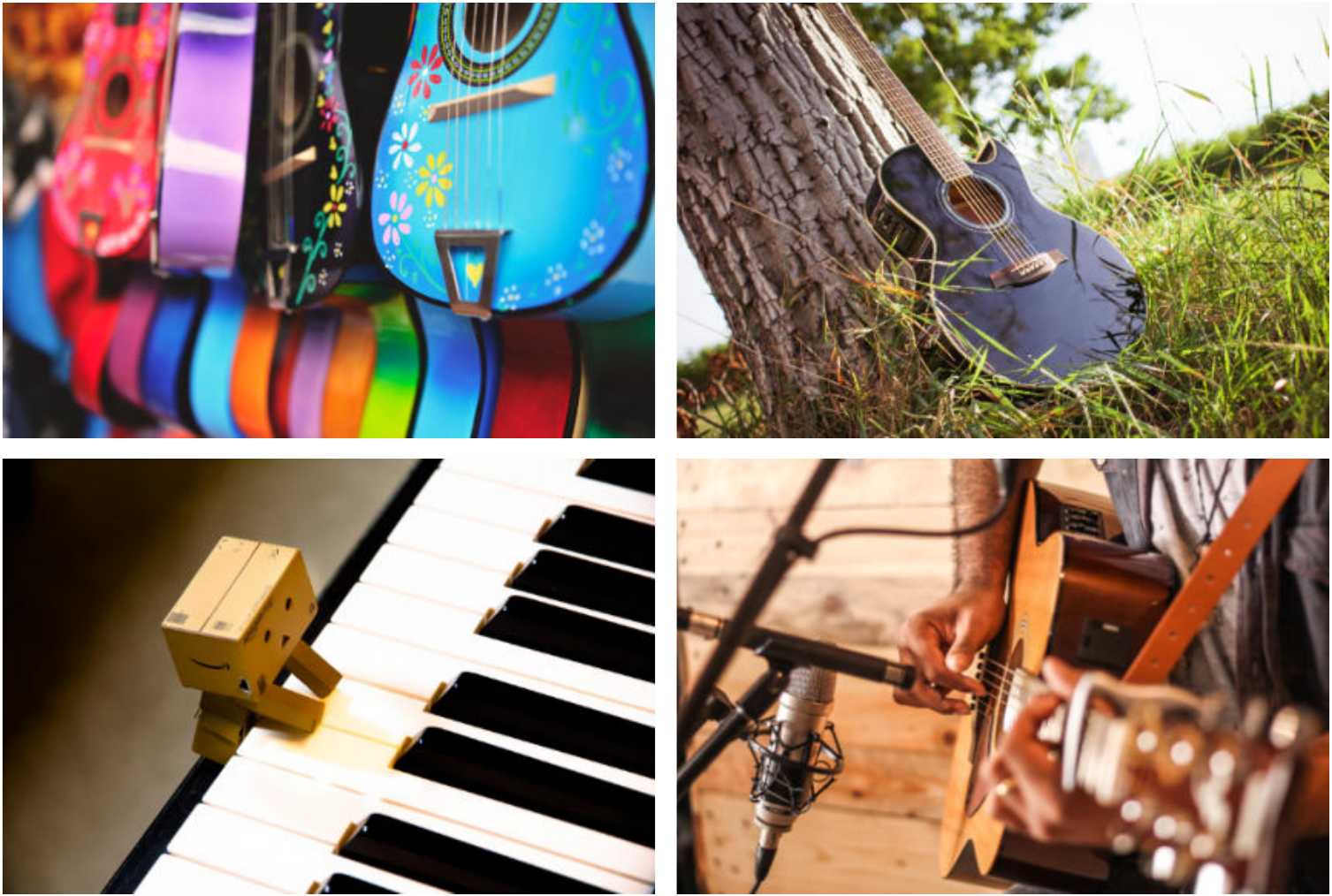}} \hskip.8em 
    \subfloat[]{\includegraphics[width=0.23\textwidth]{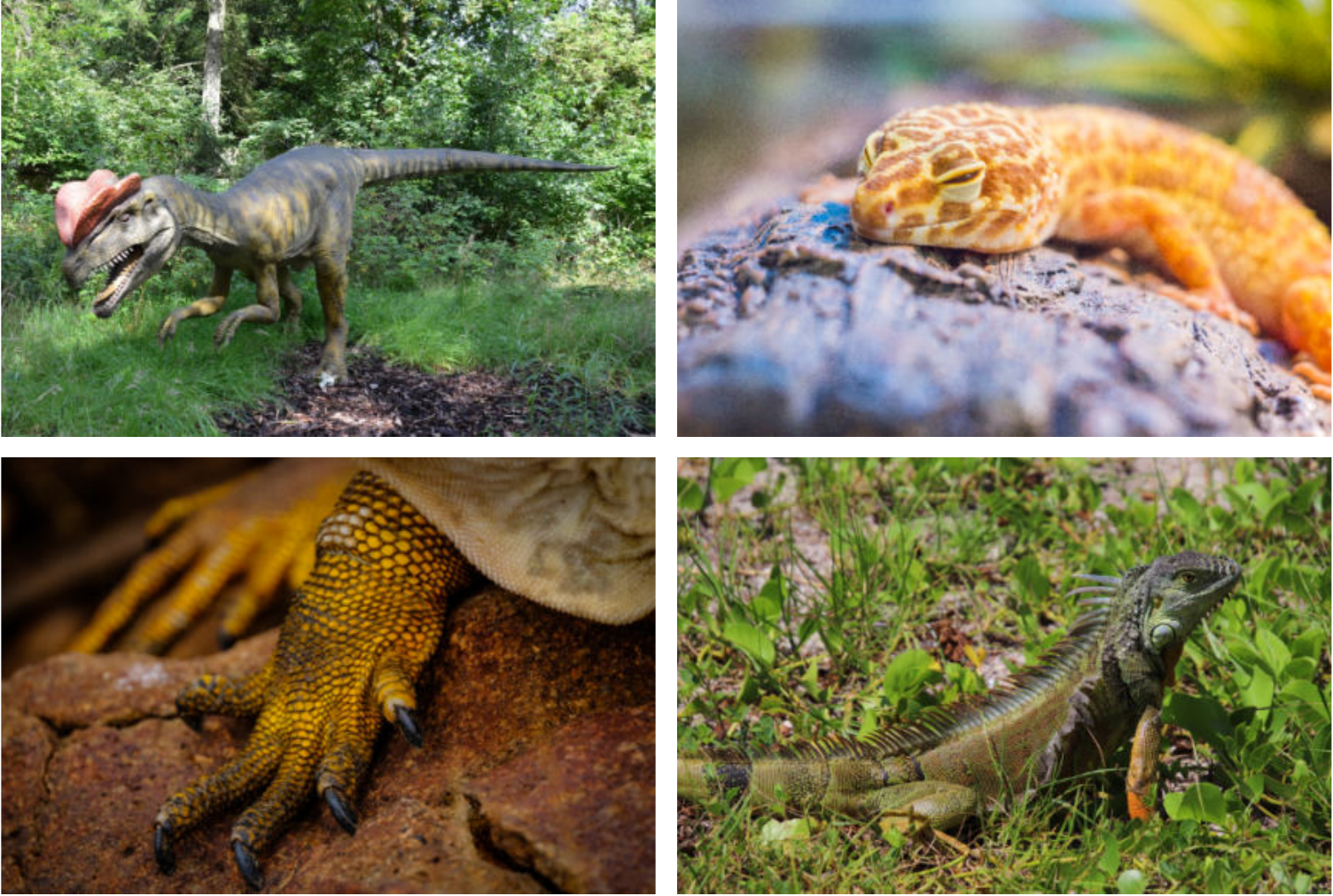}} \hskip.8em 
    \subfloat[]{\includegraphics[width=0.23\textwidth]{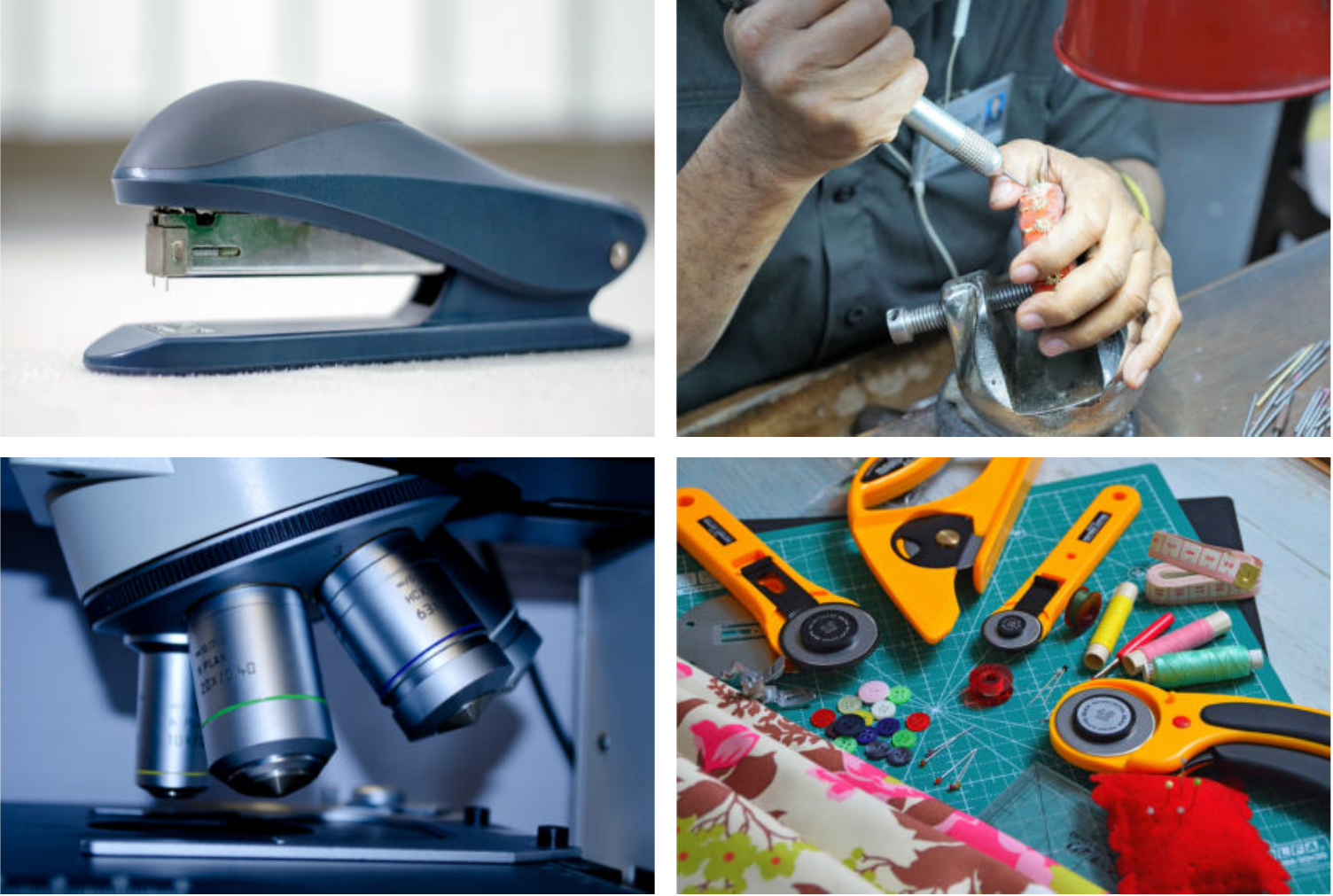}} \hskip.8em 
    \subfloat[]{\includegraphics[width=0.23\textwidth]{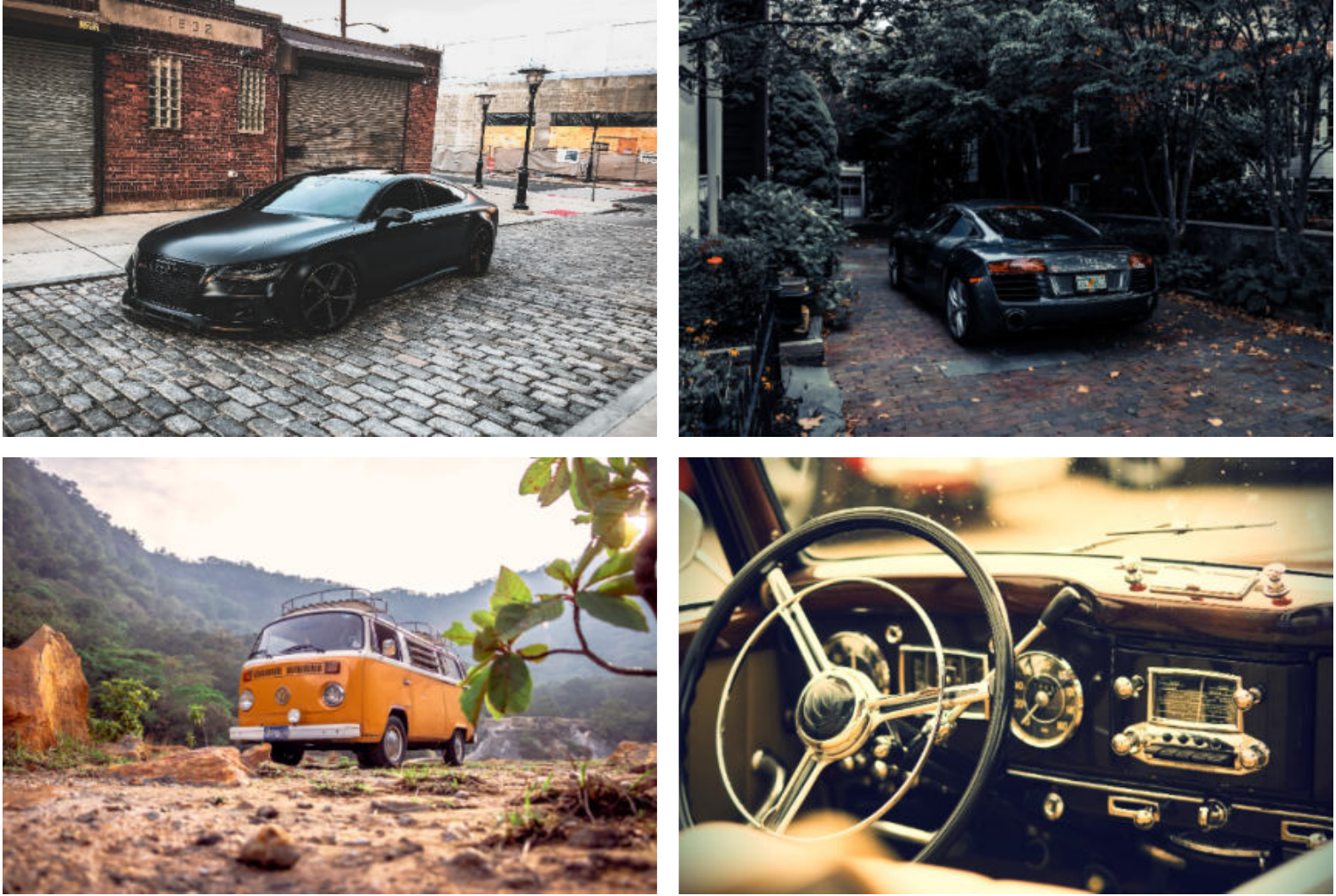}} \hskip.8em 
\end{minipage}
\caption{Sample images from the large-scale unlabeled set $\mathcal{S}$ for gMAD competition. (a) Amphibian. (b) Bird. (c) Fish. (d) Flower. (e) Fruit. (f) Furniture. (g) Geological formation. (h) Mammal. (i) Musical instrument. (j) Reptile. (k) Tool. (l) Vehicle. Images are cropped for improved visibility.}
\label{fig:datasample}
\end{figure*}
\begin{figure}[t]
\centering
   \includegraphics[width=1\linewidth]{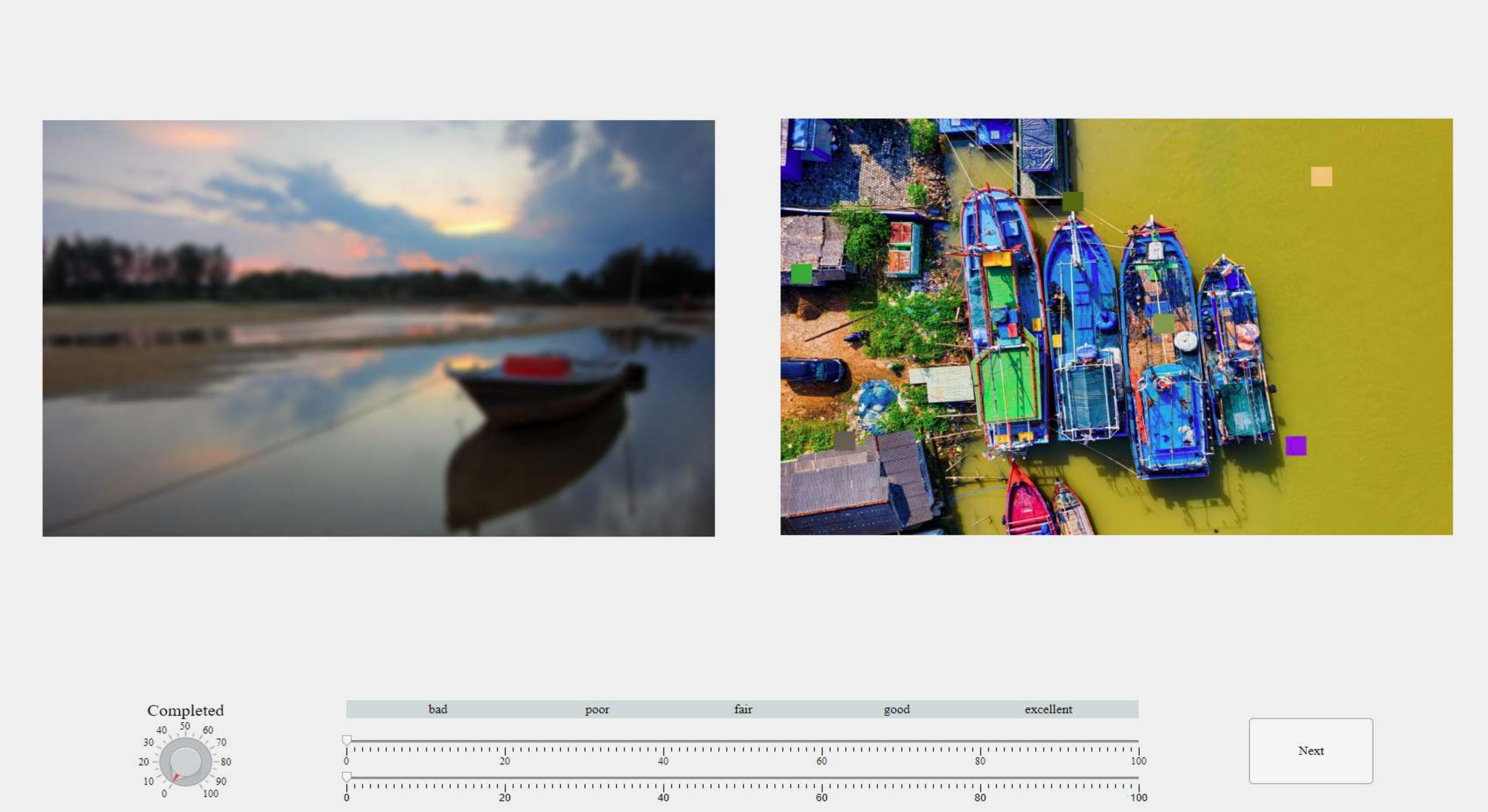}
   \caption{Graphical user interface for subjective testing.}
\label{fig:gui}
\end{figure}

\subsubsection{Construction of $\mathcal{D}_2$}We build the subject-rated image set $\mathcal{D}_2$ by combining four synthetically distorted image databases -  LIVE~\cite{sheikh2006statistical}, CSIQ~\cite{larson2010most}, TID2013~\cite{Ponomarenko201557}, and KADID-10k~\cite{lin2019kadid} (see Table~\ref{tab:dc} for details). We randomly sample $80\%$ of the reference images and their corresponding distorted ones to form  $\mathcal{D}_2$, and leave the rest for evaluation. In order to guarantee  content independence, special treatment is given when we partition overlapping reference images in LIVE and TID2013. In the end, we generate $50,000$, $50,000$, $100,000$, and $200,000$ image pairs from LIVE, CSIQ, TID2013, and KADID-10k, respectively, yielding a total of 
$400,000$.

\subsubsection{Details of Pre-Training, Fine-Tuning, and Testing}Pre-training is performed by maximizing the likelihood in Eq.~(\ref{eq:ml19}), using the Adam optimizer~\cite{kingma2014adam} with a mini-batch of $16$ and a learning rate of $10^{-4}$. After each iteration, we project the parameters $\omega$ and $\gamma$ in GDN onto the interval $[2^{-10}, \infty]$, and constrain $\gamma$ to be symmetric. The maximum epoch number is set to eight. Fine-tuning is performed by minimizing the mean fidelity loss on $\mathcal{D}_2$ in Eq.~(\ref{eq:fidelityall}). The Adam solver is adopted with a mini-batch size of $16$, a learning rate of $10^{-4}$, and a maximum epoch number of eight. During testing, we quantify the performance using the Spearman's rank correlation coefficient (SRCC) and the Pearson linear correlation coefficient
(PLCC).  For the latter, a pre-processing step is added to linearize model predictions by fitting a four-parameter monotonic function 
\begin{align}
    \hat{f}_{\hat{w}}(x) = (\eta_1 -\eta_2)/(1+\exp(-(f_{\hat{w}}(x)-\eta_3)/|\eta_4|)) + \eta_2.
    \label{eq:nm}
\end{align} 
The test set consists of four subsets of images from LIVE, CSIQ, TID2013, and KADID-10k, respectively, which we collectively denote by $\mathcal{T}$.

\subsubsection{Preliminary Results} We compare our baseline model with seven BIQA methods, including BRISQUE~\cite{mittal2012no}, NIQE~\cite{mittal2013making}, HOSA~\cite{xu2016blind}, dipIQ~\cite{ma2017dipiq}, MEON~\cite{ma2017end}, NIMA~\cite{talebi2018nima}, and deepIQA~\cite{bosse2017deep}. The former four are knowledge-driven, among which NIQE relies solely on a prior probability model of natural undistorted images and does not need MOSs for training. The latter three are data-driven DNN-based models, among which NIMA is optimized for predicting
perceptual image aesthetics using the AVA database~\cite{ava}. We also include two full-reference IQA methods - SSIM and PieAPP~\cite{prashnani2018pieapp} for reference. Table~\ref{tab:cr} shows the SRCC and PLCC results on $\mathcal{T}$ from four IQA databases. Pre-trained on $\mathcal{D}_1$, our model outperforms most BIQA models. Performance on LIVE and CSIQ is particularly strong, which is not surprising because the distortion types in the two test sets have been included during pre-training. After fine-tuning on $\mathcal{D}_2$, we observe significant performance improvements of $f_w$ on TID2013 and KADID-10k, closely matching the two full-reference models. The performance on LIVE and CSIQ drops slightly  as a consequence of balancing  more distortion types.  In summary, by combining the training techniques in~\cite{ma2019blind} and~\cite{zhang2019learning}, we arrive at a top-performing BIQA model that is capable of handling a number of synthetic distortions.

\subsection{Specification of the Active Fine-Tuning Cycle}
\subsubsection{Construction of $\mathcal{S}$}
We collect a large-scale unlabeled image set $\mathcal{S}$ as the candidate  pool  to seek gMAD examples for active fine-tuning. Specifically, we first download high-quality and high-definition natural images from the Internet that carry  Creative Common licenses. They can be loosely grouped into twelve categories: amphibian, bird, fish, flower, fruit, furniture, geological formation, mammal, musical instrument, reptile, tool, and vehicle (see representative images in Fig.~\ref{fig:datasample}). We remove near-duplicate images using the command line tool $\mathrm{imgdupes}()$\footnote{https://github.com/knjcode/imgdupes\#against-large-dataset}, and delete those with inappropriate content. This leaves us $10,000$ natural photographic images, and  the number in each category is approximately the same. We downsample the images to a maximum width or height of $1,024$ as a way of further reducing possibly visible artifacts. After data screening, we add $25$ types of distortions with five levels of severity, which are the same in KADID-10k~\cite{lin2019kadid} and can be roughly classified into seven categories: blurring,  color-related distortion, compression, noise-related distortion, intensity change, contrast change and others. Finally, for each reference image, we randomly choose 5 out of 25 distortion types and 2 out of  5 levels, resulting in a total of $5\times 2\times 10,000=100,000$ distorted images.

\subsubsection{Construction of $\mathcal{U}^{(t)}$}We let our method compete with nine state-of-the-art full-reference IQA models - SSIM~\cite{wang2004image}, MS-SSIM~\cite{wang2003multiscale}, NLPD~\cite{laparra2016perceptual}, VSI~\cite{zhang2014vsi},  MAD~\cite{larson2010most}, VIF~\cite{sheikh2006image}, MDSI~\cite{nafchi2016mean}, PieAPP~\cite{prashnani2018pieapp}, and WaDIQaM~\cite{bosse2017deep}, among which the former seven  are knowledge-driven, while the latter two  are purely data-driven methods based on DNNs. All implementations are obtained from the original authors, except for WaDIQaM which we use a publicly available re-implementation\footnote{https://github.com/lidq92/WaDIQaM}. gMAD requires all competing models to work in the same perceptual scale. Therefore, we map all model predictions using Eq.~(\ref{eq:nm}) onto the LIVE MOS scale $[0,100]$, with higher values indicating better perceptual quality.  Five levels ($l=5$) are specified to roughly cover  bad, poor, fair, good, and excellent quality. The quality range (\ie, bin width) is half of the mean std in LIVE, ensuring that the images in the same level have similar quality in terms of the
defender model. Two types of gMAD pairs are queried by treating our baseline model as the defender and the attacker, respectively. We take the subjective testing effort into account, and search for a maximum of $k=12$ pairs at each quality level. During this process, we find that if our model fails in one corner case, more failure examples of the same case may be picked out repeatedly by other competing models. To enhance content and distortion  diversity of the selected images, we enforce several additional constraints on each pairwise model comparison: (1) images of the same content appear at most twice; (2) images of the same distortion type appear at most three times; (3) combinations of the same two distortion types appear at most once.

\subsubsection{Subjective Testing}
\label{sec:subjective}
We set up the subjective experiment in an office environment with a normal indoor illumination level. The display we use is a true-color LED monitor with the resolution of $2,560\times1,920$ pixels, and we calibrate it according to the recommendation of ITU-R BT.500~\cite{video2000final}. 
Fig.~\ref{fig:gui} illustrates the graphical user interface we customize for this experiment. A gMAD pair is rendered at full image resolution, but in random spatial order. Two scale-and-slider applets are utilized to collect the quality score of each image, with $0$ 
and $100$ indicating worst and best quality, respectively. The viewing distance is fixed to $32$ pixels per degree of visual angle. For each $\mathcal{U}^{(t)}$, we gather data from fifteen subjects with normal or correct-to-normal visual acuity. They have general knowledge of image processing and computer vision, but do not know the detailed purpose of the study. We include a training session to familiarize them with image distortions. Each subject is asked to give scores to all gMAD images. To minimize the influence of the fatigue effect, the subjects are allowed to take a break after a maximum of 30-minute experiment. We process the raw data using the outlier detection and subject rejection algorithm in~\cite{bt2002methodology}. In total, we perform three rounds of subjective experiments ($T=3$). $\mathcal{L}^{(1)}$ and $\mathcal{L}^{(2)}$ are used to evaluate and  refine $f_w$ in the active fine-tuning cycle, while $\mathcal{L}^{(3)}$ is reserved for testing. After data purification, we find that all subjects are valid, and  $2.82\%$, $2.68\%$ and  $2.26\%$ of all ratings are identified as outliers and subsequently removed in $\mathcal{L}^{(1)}$, $\mathcal{L}^{(2)}$ and $\mathcal{L}^{(3)}$, respectively.

\begin{figure}[t]
  \centering
  \subfloat[]{\includegraphics[width=0.5\textwidth]{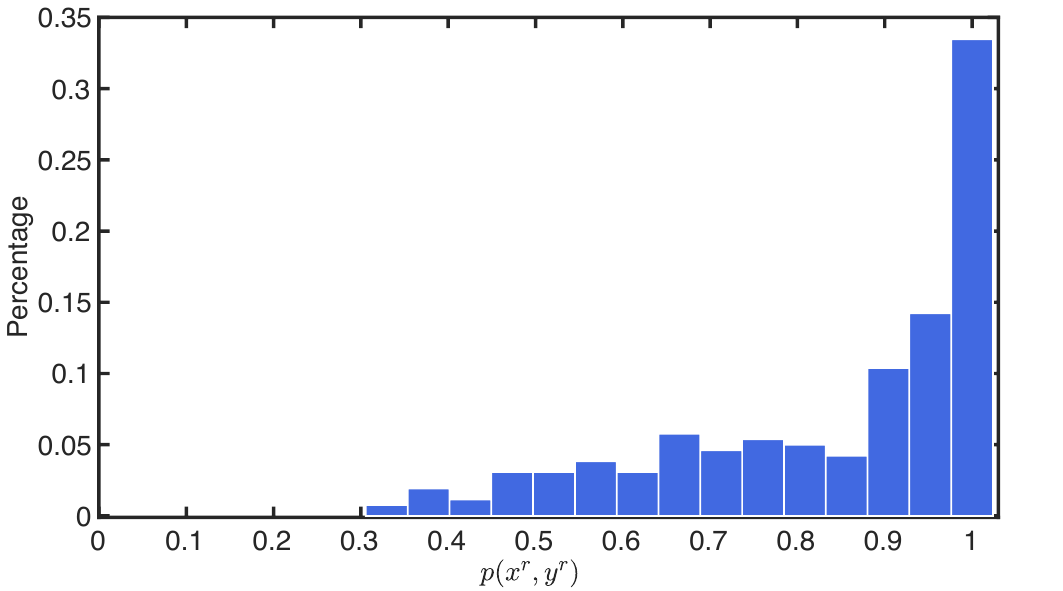}}\hskip.2em
  \subfloat[]{\includegraphics[width=0.5\textwidth]{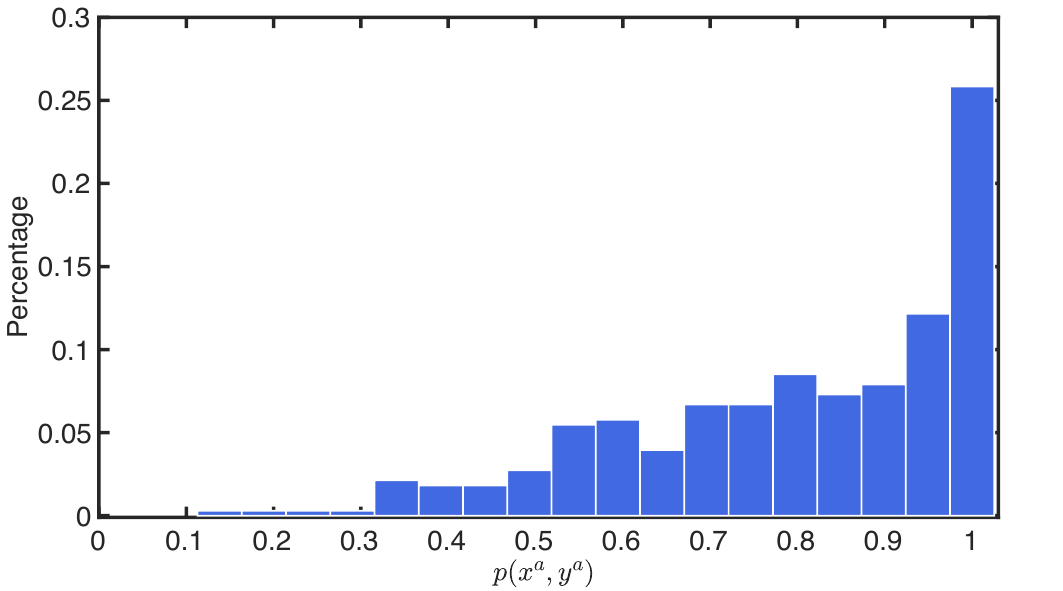}}
  \caption{The empirical distributions of (a) $p(x^r, y^r)$ and (b) $p(x^a, y^a)$ on $\mathcal{L}^{(1)}$. It is clear that full-reference IQA methods (as attackers) can easily falsify our BIQA model, and vice versa.}
  \label{fig:probability distribution}
\end{figure}

Fig.~\ref{fig:probability distribution} shows the empirical distributions of $p(x^r, y^r)$ and $p(x^a, y^a)$  computed by Eq.~(\ref{eq:gt}) on $\mathcal{L}^{(1)}$. When the baseline model is the defender, it is effortless for the set of full-reference IQA methods to spot its failures, as evidenced by a large percentage of pairs with $p(x^r, y^r) >0.8$ (belonging to Case I). These are strong counterexamples of $f_w$, which are informative in active fine-tuning. When our model works as the attacker, it performs surprisingly well in falsifying full-reference models with a large portion of the selected pairs belonging to Case IV. This adds new direct evidence to our claim of the top performance of the baseline model before active fine-tuning.

\begin{table}
\caption{Correlation (SRCC and PLCC) results on the gMAD image sets. Our results on $\mathcal{L}^{(1)}$, $\mathcal{L}^{(2)}$, $\mathcal{L}^{(3)}$ are obtained by the proposed method before active fine-tuning, after the first round of active fine-tuning on $\mathcal{L}^{(1)}$, and after the second round of active fine-tuning on both $\mathcal{L}^{(1)}$ and $\mathcal{L}^{(2)}$, respectively. See Algorithm~\ref{alg:af} for the detailed procedure}
\centering
\begin{threeparttable}
\begin{tabular}{l|cccc}
\toprule
SRCC & KADID-10k & $\mathcal{L}^{(1)}$ & $\mathcal{L}^{(2)}$ & $\mathcal{L}^{(3)}$ \\
\hline
SSIM~\cite{wang2004image} & $0.752$ & $0.615$ & $0.482$ & $0.499$\\
MS-SSIM~\cite{wang2003multiscale} & $0.826$ & $0.745$ & $0.616$ & $0.652$\\
NLPD~\cite{laparra2016perceptual} & $0.812$ & $0.767$ & $0.624$ & $0.646$\\
VSI~\cite{zhang2014vsi} & $0.879$ & $0.772$ & $0.685$ & $0.697$\\
MAD~\cite{larson2010most} & $0.799$ & $0.731$ & $0.648$ & $0.638$\\
VIF~\cite{sheikh2006image} & $0.679$ & $0.721$ & $0.677$ & $0.679$\\
MDSI~\cite{nafchi2016mean} & $\bf 0.887$ & $0.759$ & $0.669$ & $0.694$\\
PieAPP~\cite{prashnani2018pieapp} & $0.865$ & $\bf 0.783$ & $0.718$ & $0.761$\\
WaDIQaM~\cite{bosse2017deep} & $\bf 0.966$\tnote{*} & $\bf 0.814$ & $\bf 0.730$ & $\bf 0.773$\\
\hline
Ours & $-$ & $0.633$ & $\bf 0.818$ & $\bf 0.813$\\
\midrule
\midrule
PLCC & KADID-10k &  $\mathcal{L}^{(1)}$ & $\mathcal{L}^{(2)}$ & $\mathcal{L}^{(3)}$ \\
\hline
SSIM & $0.743$ & $0.659$ & $0.484$ & $0.514$\\
MS-SSIM & $0.820$ & $0.739$ & $0.603$ & $0.644$\\
NLPD & $0.811$ & $0.773$ & $0.629$ & $0.646$\\
VSI & $0.877$ & $0.774$ & $0.674$ & $0.688$\\
MAD & $0.799$ & $0.736$ & $0.649$ & $0.640$\\
VIF & $0.686$ & $0.760$ & $0.698$ & $0.700$\\
MDSI & $\bf 0.887$ & $0.776$ & $0.669$ & $0.689$\\
PieAPP & $0.866$ & $\bf 0.800$ & $0.722$ & $0.765$\\
WaDIQaM & $\bf 0.967$\tnote{*} & $\bf 0.818$ & $\bf 0.732$ & $\bf 0.770$\\
\hline
Ours & $-$ & $0.630$ & $\bf 0.823$ & $\bf 0.828$\\
\bottomrule
\end{tabular}
\begin{tablenotes}
        \footnotesize
        \item[*] WaDIQaM is trained on KADID-10k.
      \end{tablenotes}
\end{threeparttable}
\label{tab:mainresult}
\end{table}

\begin{figure}[t]
\centering
\includegraphics[width=1\linewidth]{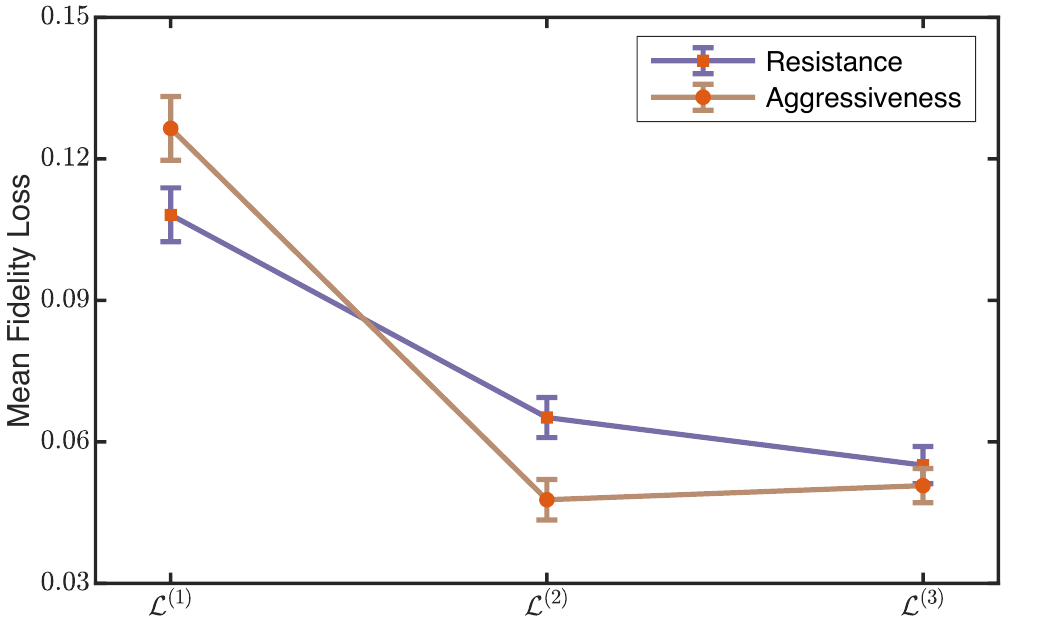}
\caption{The progress of our method in terms of the mean fidelity loss ($\pm$ standard error) on the gMAD sets, when playing the role of the defender and the attacker, respectively.}
\label{fig:mean fidelity}
\end{figure}
\subsubsection{Details of Active Fine-Tuning}
For each round of active fine-tuning, we minimize the weighted mean fidelity loss in Eq.~(\ref{eq:af}). The Adam optimizer is used with a mini-batch size of $16$ - half from $\mathcal{D}_2$ and half from $\mathcal{D}_3'$. This amounts to oversampling $\mathcal{D}_3'$, and provides an equivalent implementation of  Eq.~(\ref{eq:af}) in the mini-batch setting. The learning rates for shallow layers (up to the second GDN layer) and deep layers are set to $10^{-5}$ and $10^{-4}$, respectively. The maximum epoch number is set to eight. SRCC, PLCC, and the mean fidelity loss are used to quantify the performance during testing.

\begin{figure*}[t]
\centering
\subfloat[]{\includegraphics[width=0.27\textwidth]{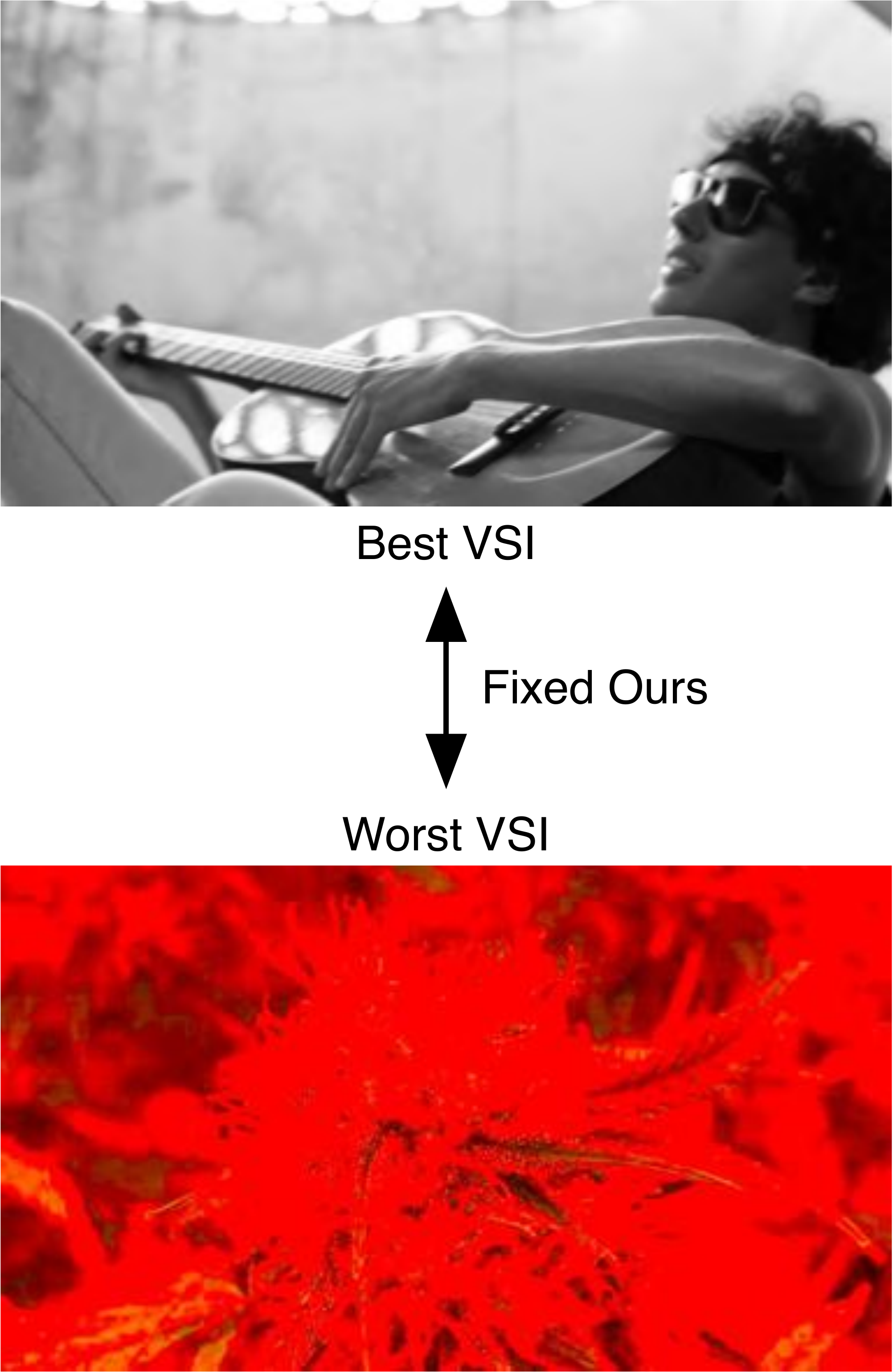}}\hskip.5em\label{fig:vsi as attacker(a)}
\subfloat[]{\includegraphics[width=0.27\textwidth]{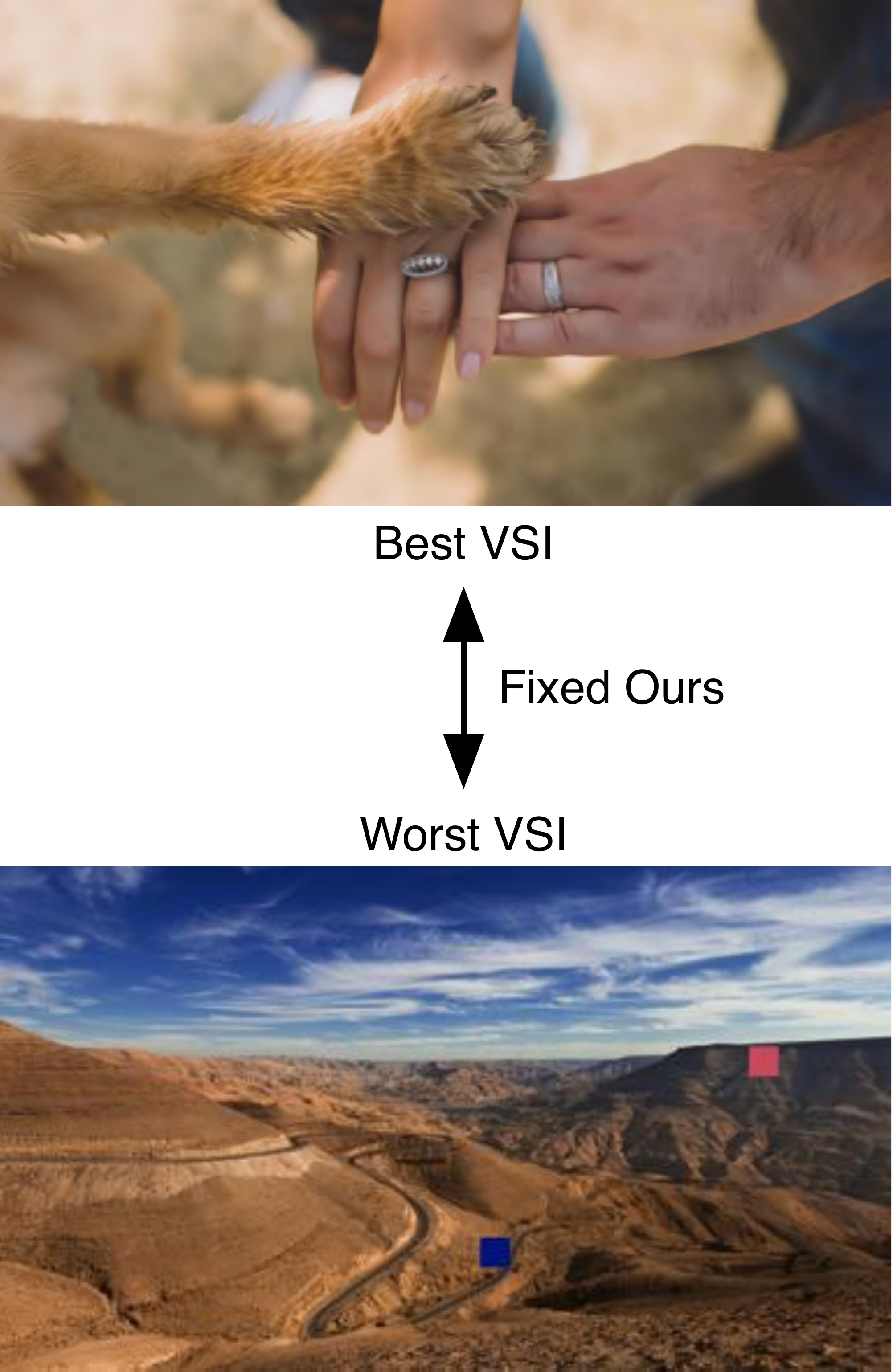}}\hskip.5em
\subfloat[]{\includegraphics[width=0.27\textwidth]{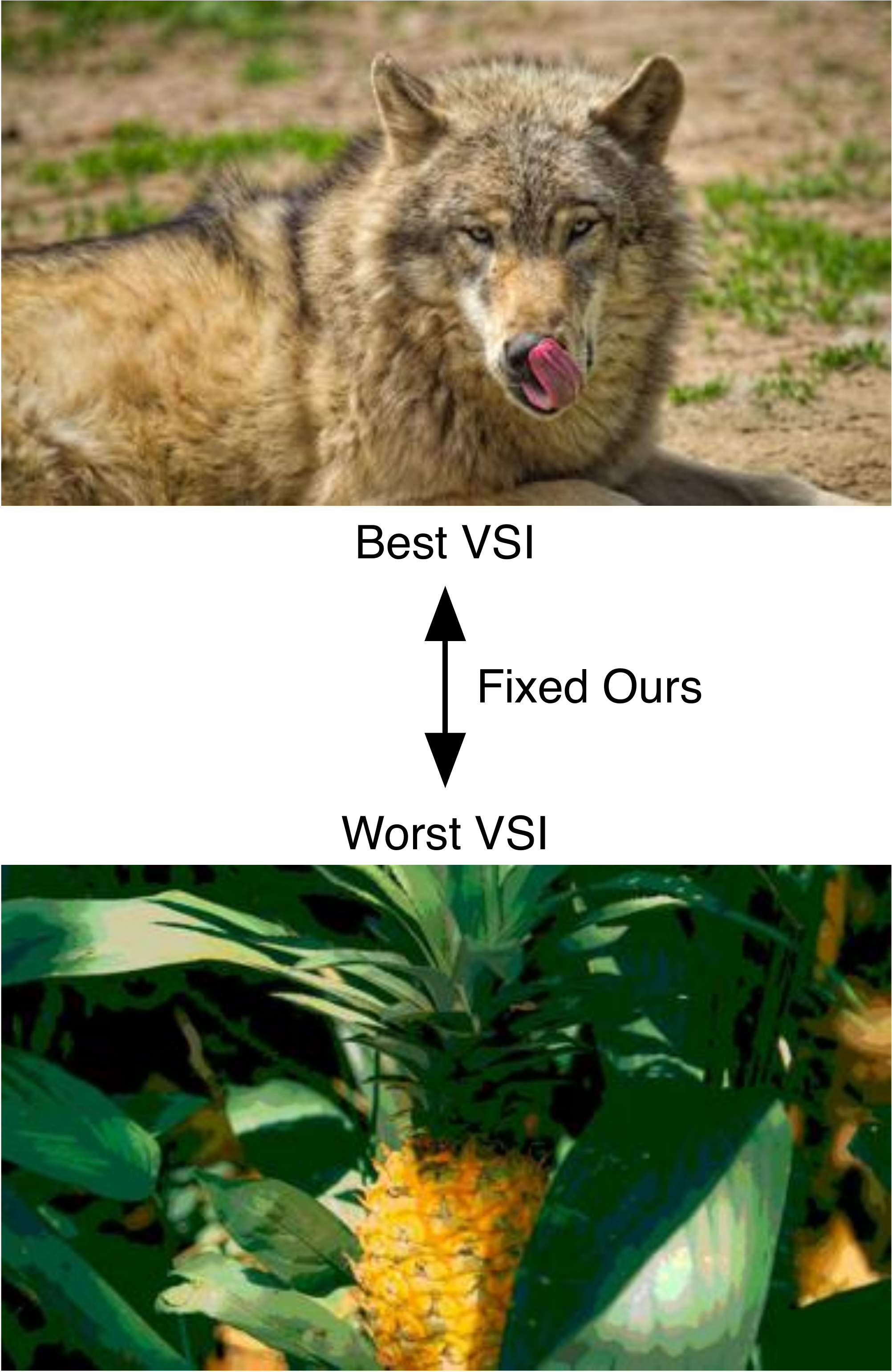}}\hskip.5em \\
\caption{gMAD image pairs with the maximum fidelity losses (\ie, the worst-case samples) selected in (a) $\mathcal{L}^{(1)}$, (b) $\mathcal{L}^{(2)}$, and (c) $\mathcal{L}^{(3)}$, respectively, when our model is the defender and VSI~\cite{zhang2014vsi} is the attacker.}
\label{fig:vsi as attacker}
\end{figure*}

\begin{figure*}[t]
\centering
\subfloat[]{\includegraphics[width=0.27\textwidth]{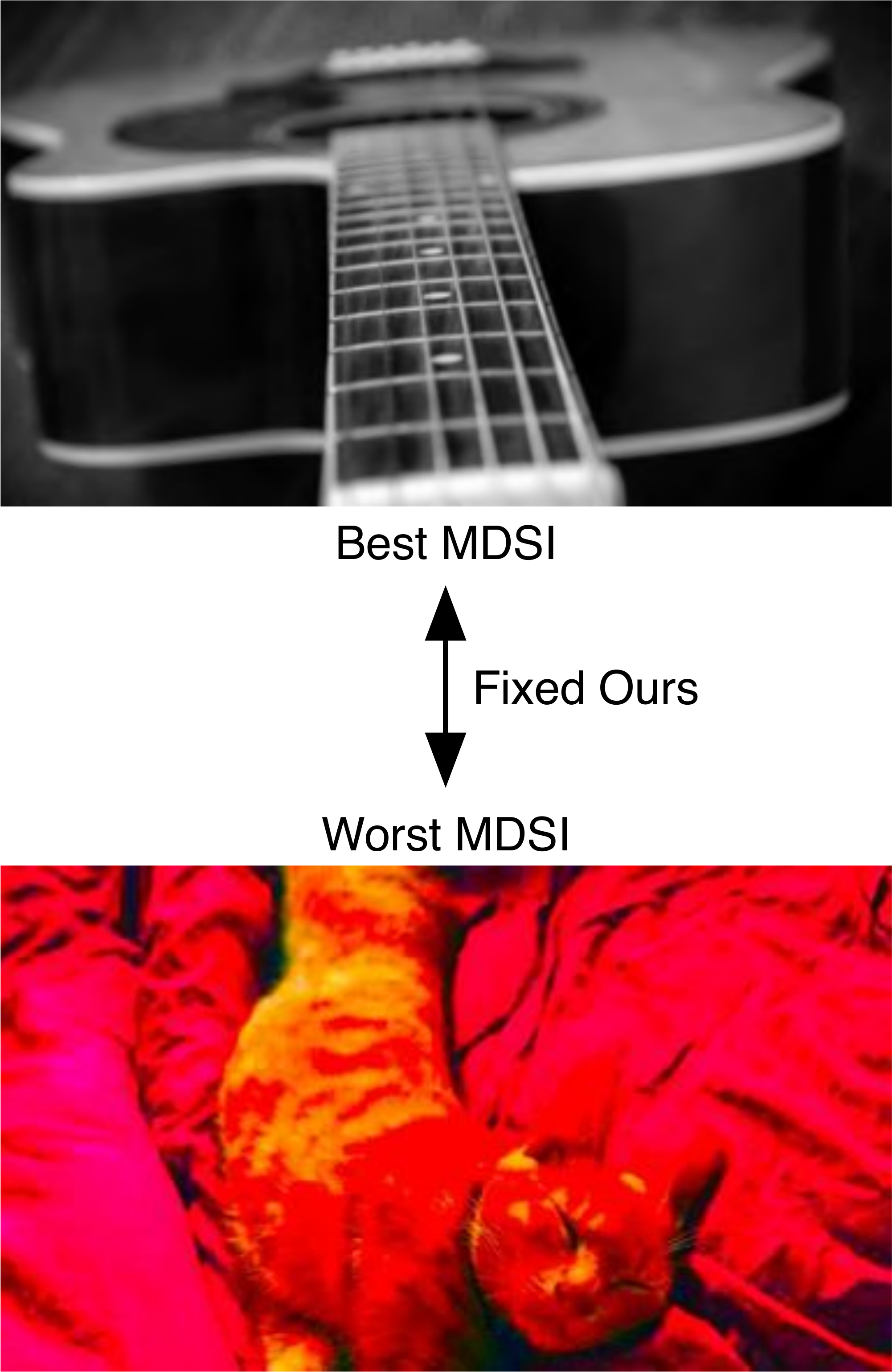}}\hskip.5em
\subfloat[]{\includegraphics[width=0.27\textwidth]{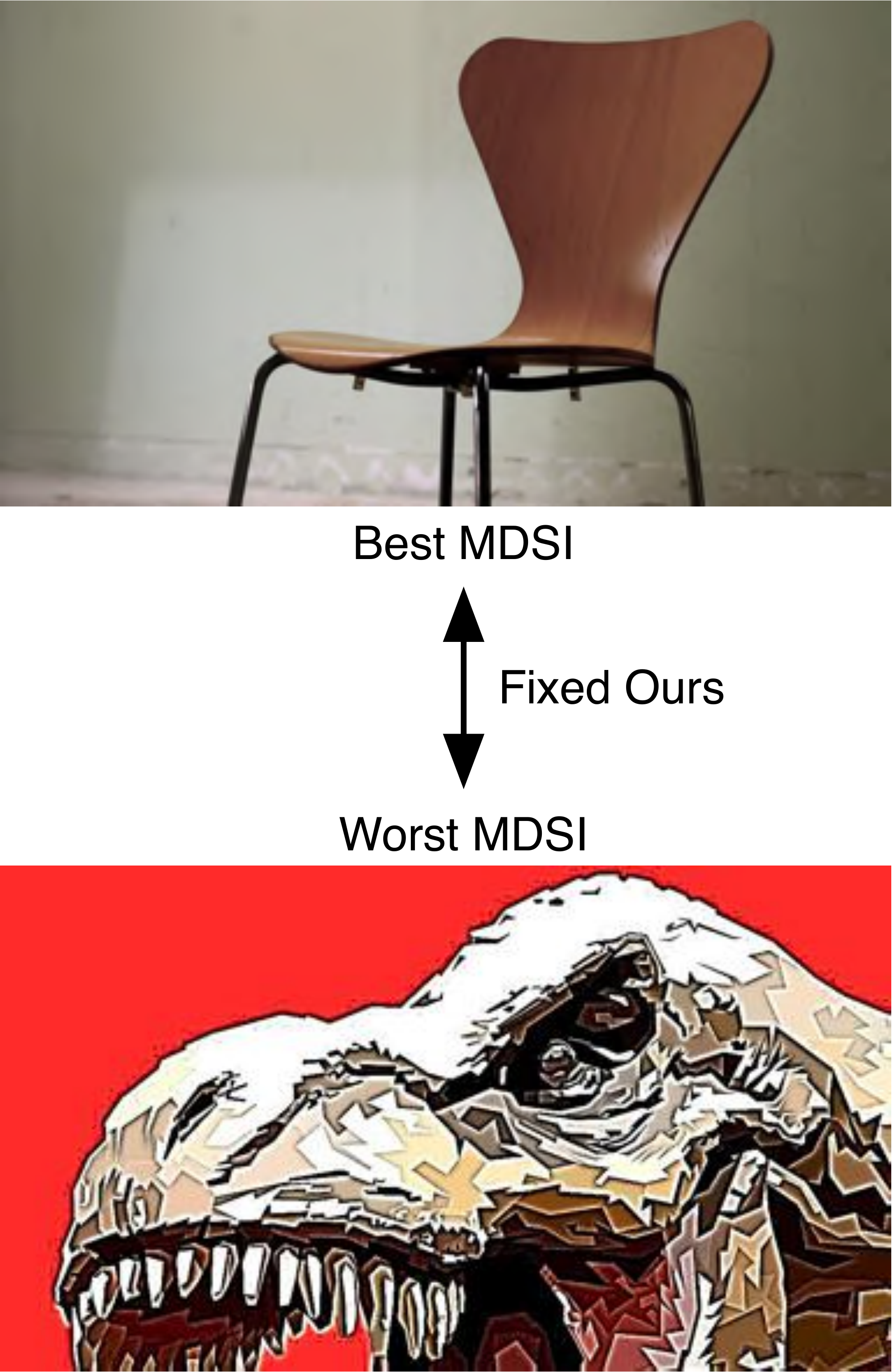}}\hskip.5em
\subfloat[]{\includegraphics[width=0.27\textwidth]{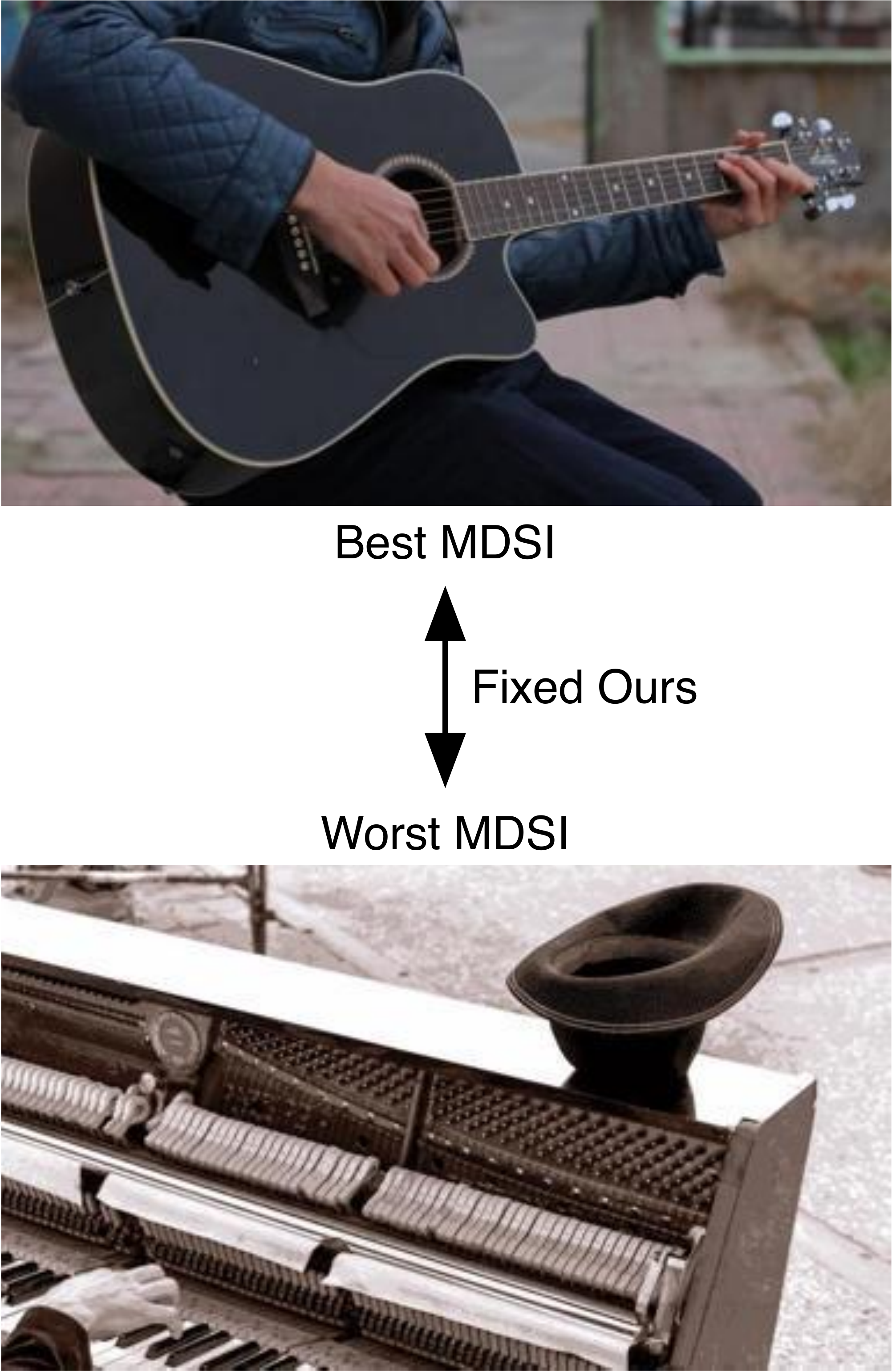}}\hskip.5em \\
\caption{gMAD image pairs with the maximum fidelity losses selected in (a) $\mathcal{L}^{(1)}$, (b) $\mathcal{L}^{(2)}$, and (c) $\mathcal{L}^{(3)}$, respectively, when our model is the defender and MDSI~\cite{nafchi2016mean} is the attacker.}
\label{fig:MDSI as attacker}
\end{figure*} 

\begin{figure*}[t]
\centering
\subfloat[]{\includegraphics[width=0.27\textwidth]{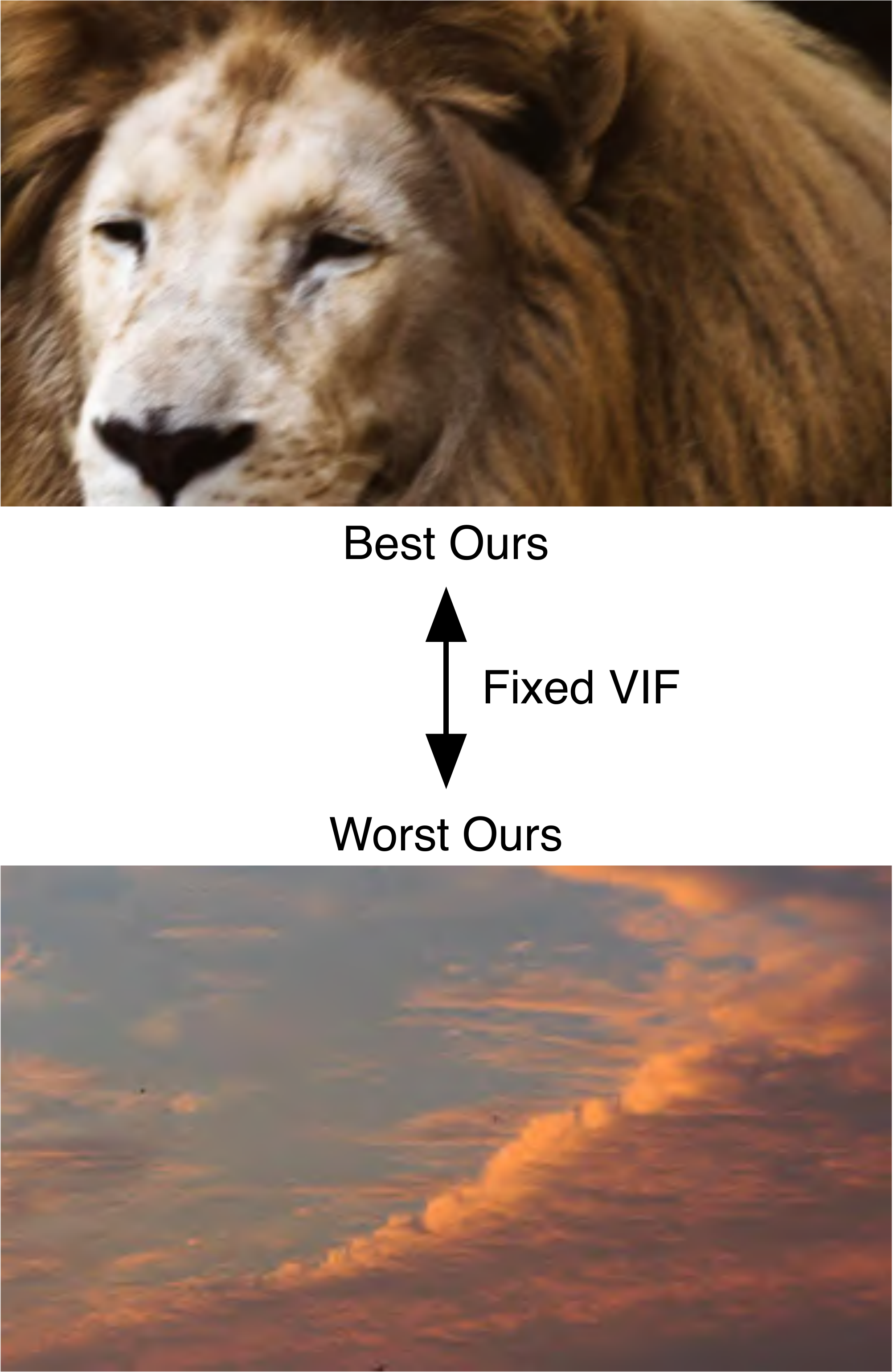}}\hskip.5em
\subfloat[]{\includegraphics[width=0.27\textwidth]{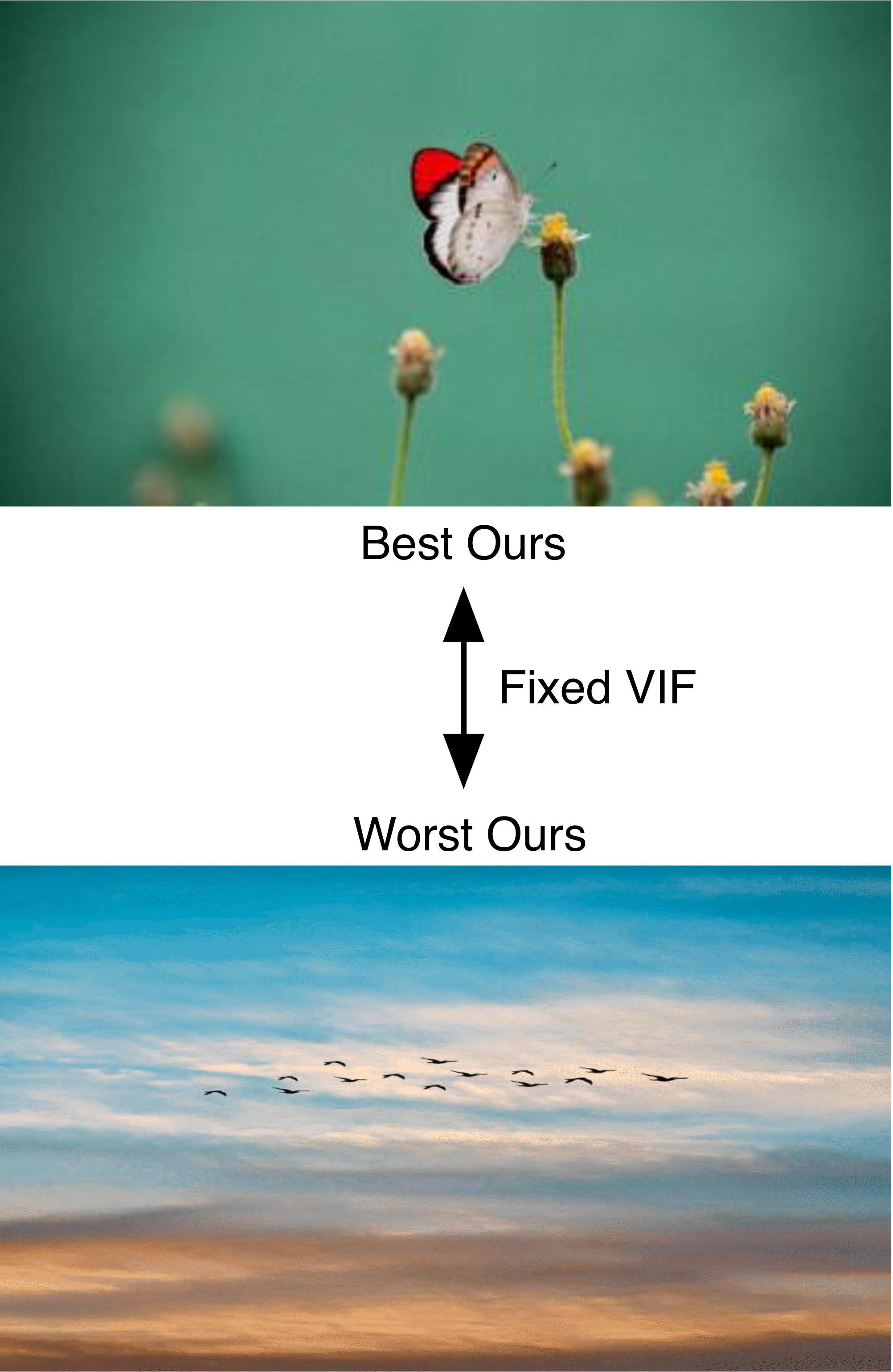}}\hskip.5em
\subfloat[]{\includegraphics[width=0.27\textwidth]{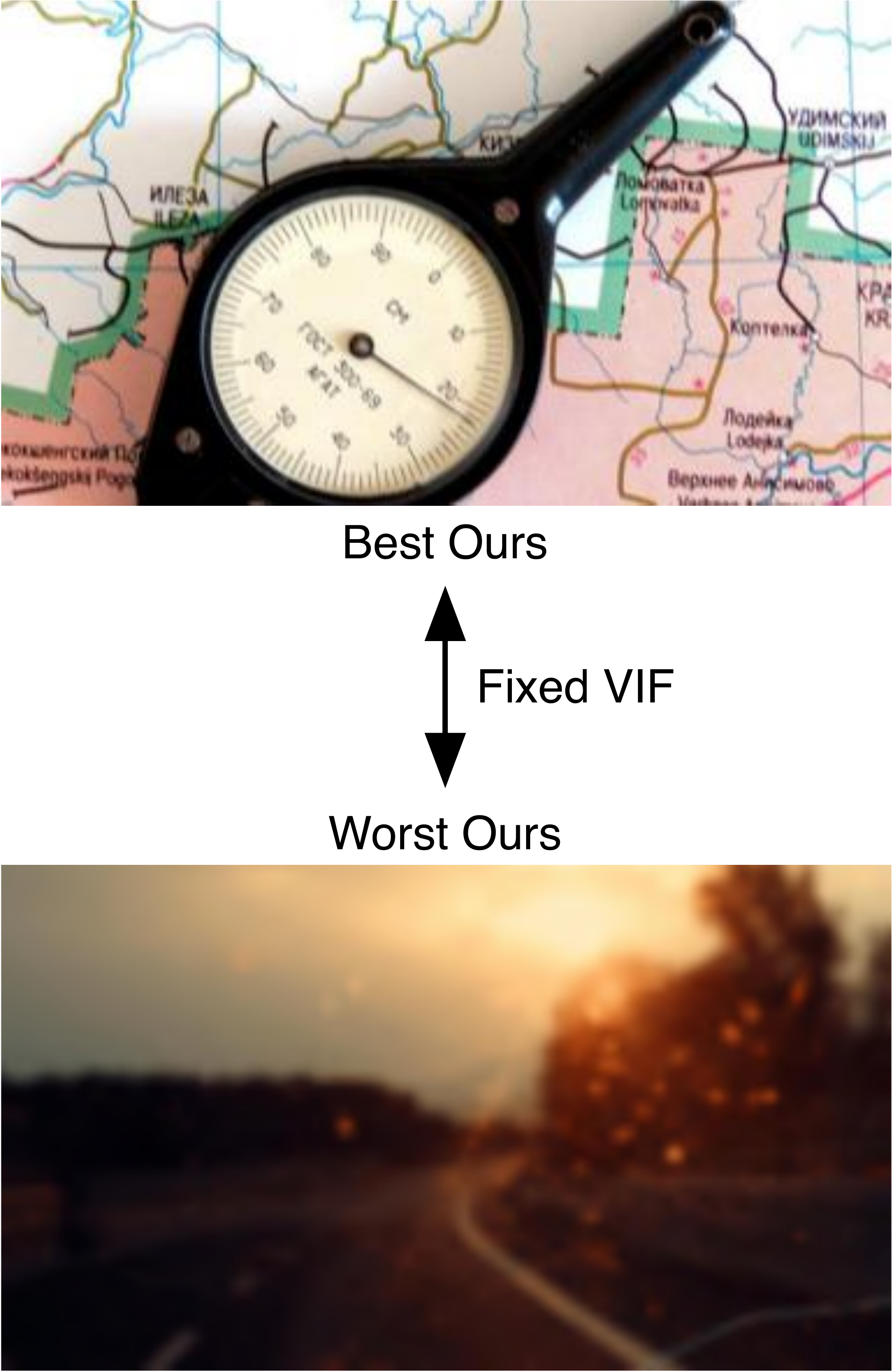}}\hskip.5em \\
\caption{gMAD image pairs with the maximum fidelity losses (\ie, the worst-case samples) selected in (a) $\mathcal{L}^{(1)}$, (b) $\mathcal{L}^{(2)}$, and (c) $\mathcal{L}^{(3)}$, respectively, when VIF~\cite{sheikh2006image} is the defender and our model is the attacker.}
\label{fig:vif as defender}
\end{figure*} 

\begin{figure*}[t]
\centering
\subfloat[]{\includegraphics[width=0.27\textwidth]{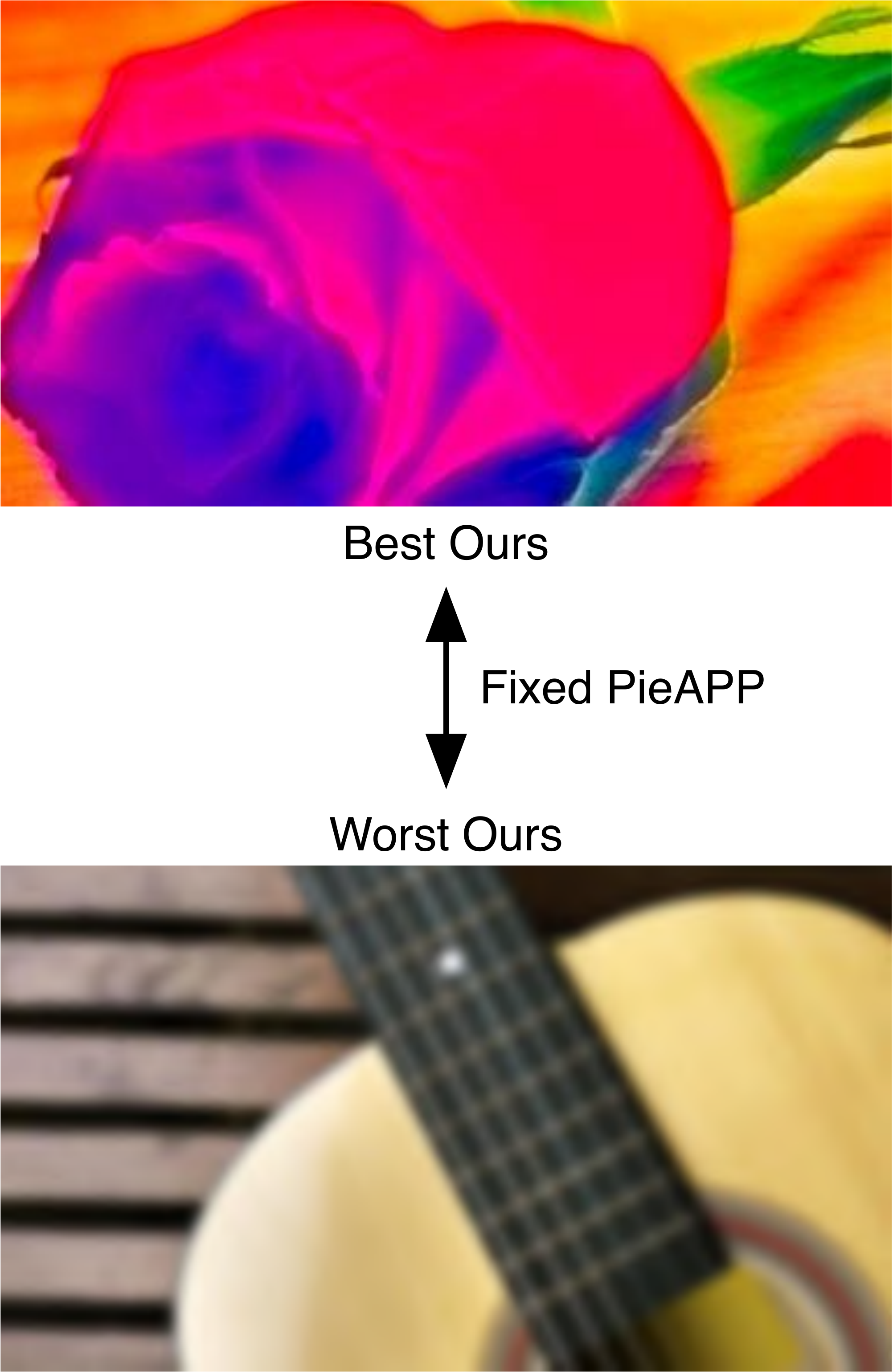}}\hskip.5em
\subfloat[]{\includegraphics[width=0.27\textwidth]{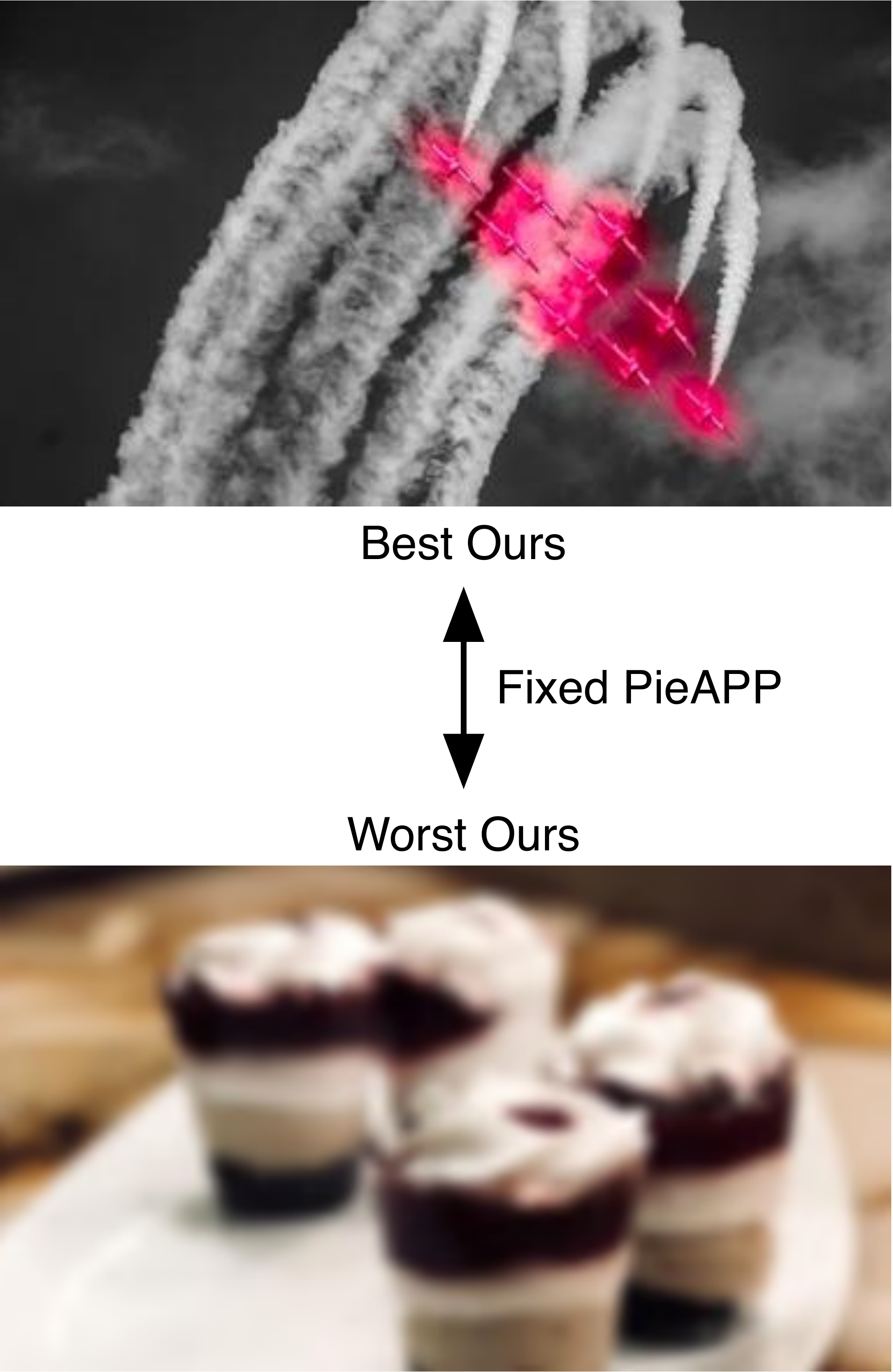}}\hskip.5em
\subfloat[]{\includegraphics[width=0.27\textwidth]{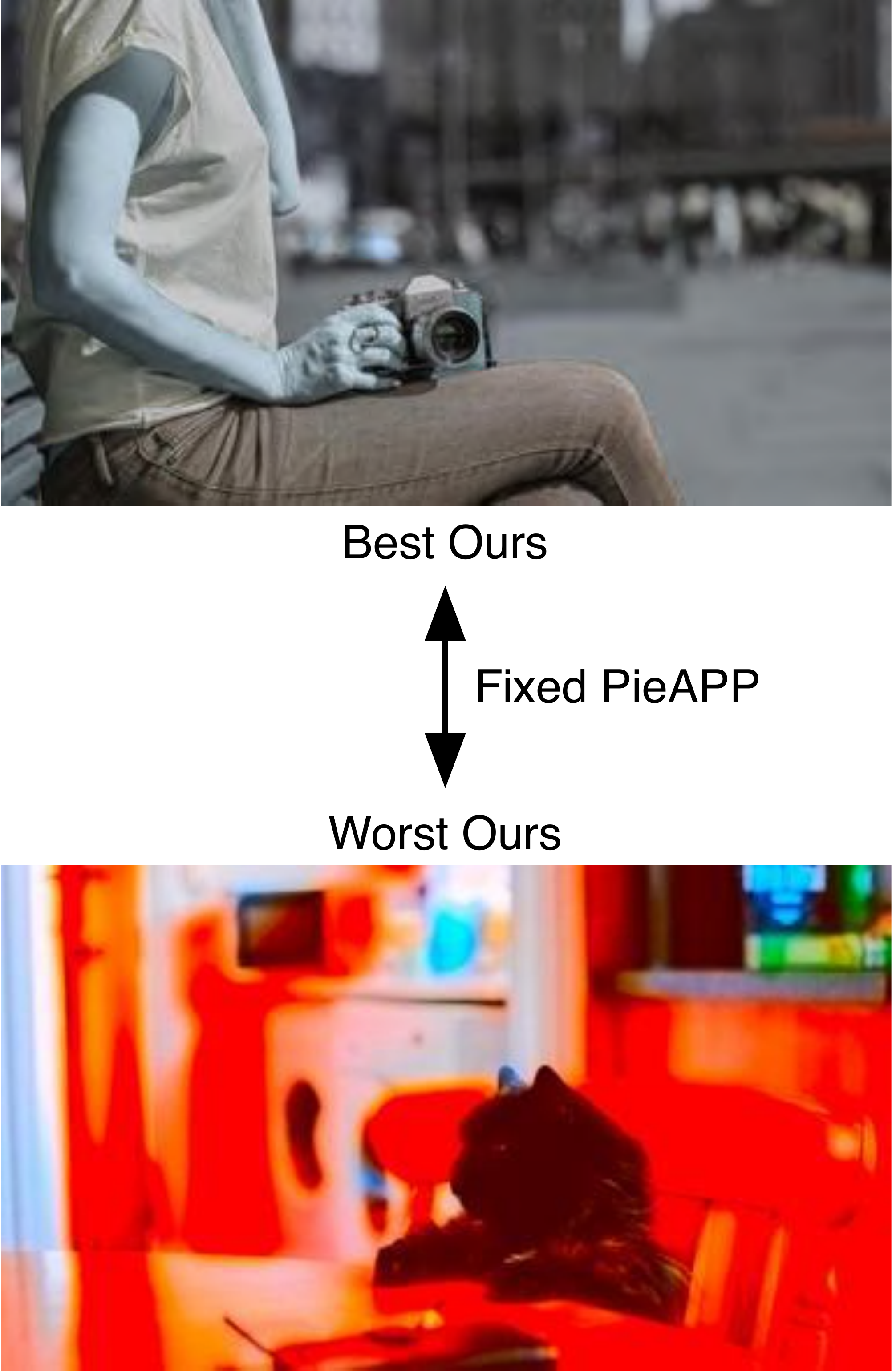}}\hskip.5em \\
\caption{gMAD image pairs with the maximum fidelity losses selected in (a) $\mathcal{L}^{(1)}$, (b) $\mathcal{L}^{(2)}$, and (c) $\mathcal{L}^{(3)}$, respectively, when PieAPP~\cite{prashnani2018pieapp} is the defender and our model is the attacker.}
\label{fig:PieAPP as defender}
\end{figure*} 

\begin{figure*}[t]
\centering
\begin{minipage}[t]{1.0\textwidth}
    \centering
    \subfloat[]{\includegraphics[width=0.27\textwidth]{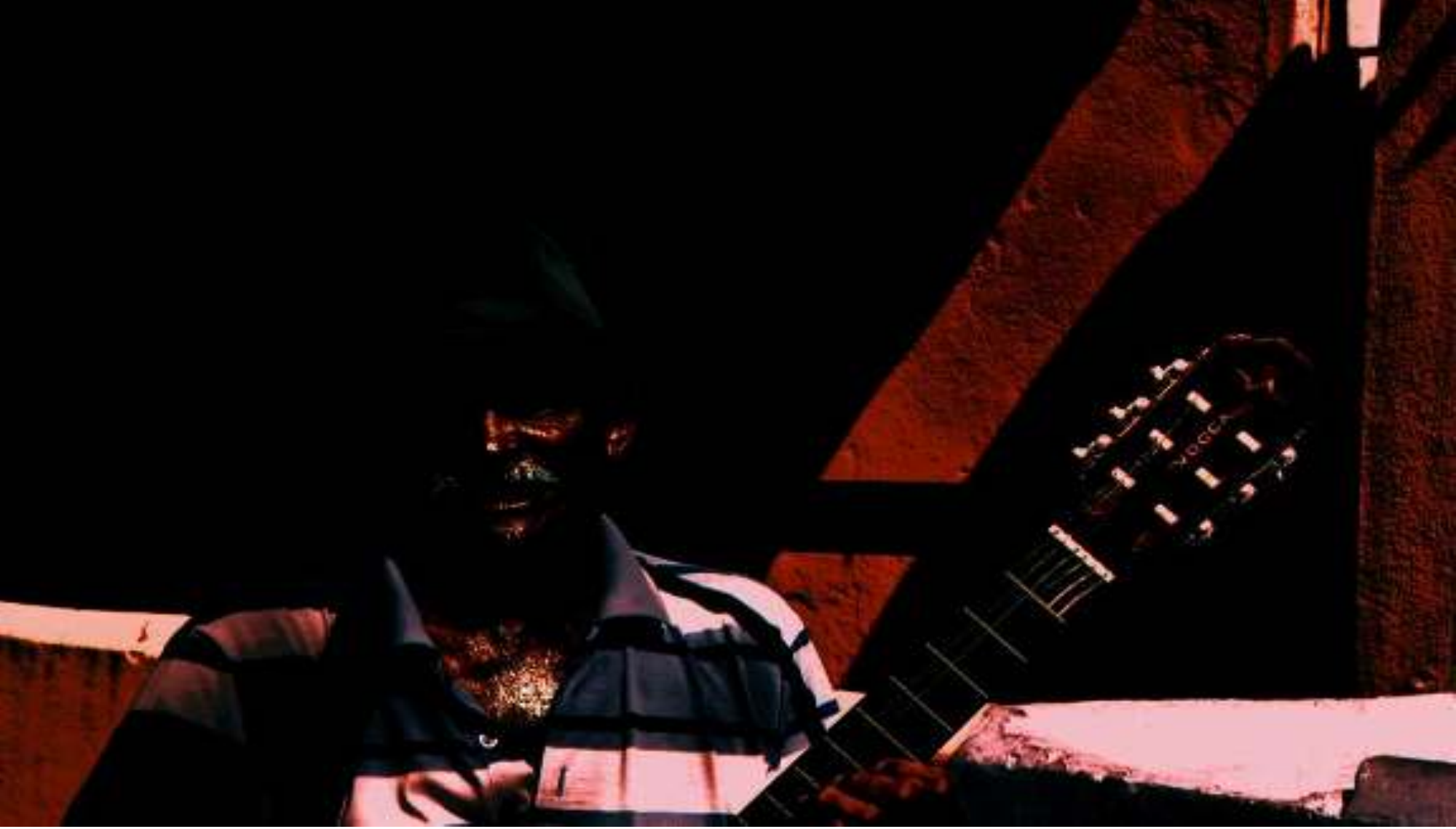}}\hskip.5em
    \subfloat[]{\includegraphics[width=0.27\textwidth]{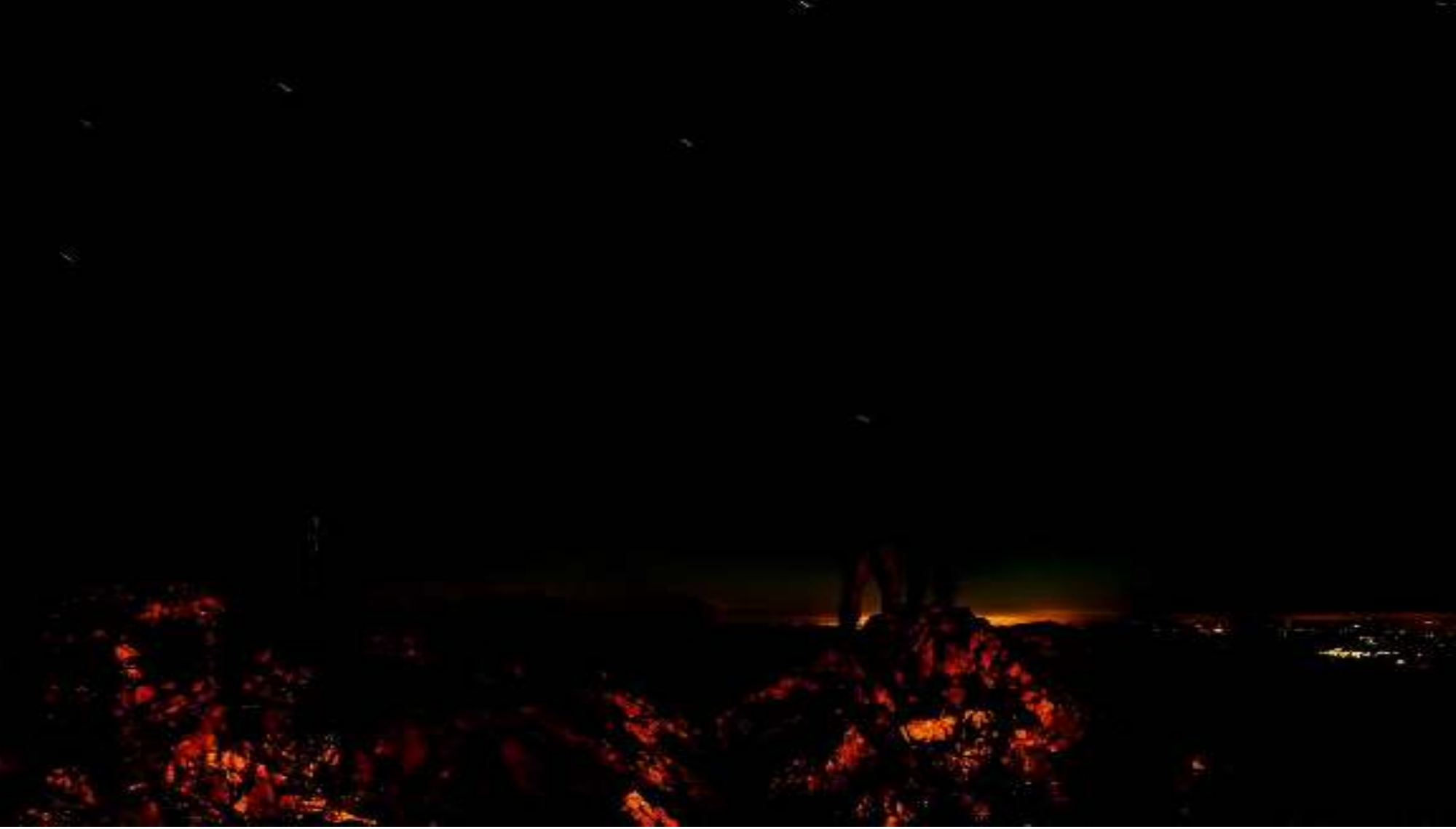}}\hskip.5em
    \subfloat[]{\includegraphics[width=0.27\textwidth]{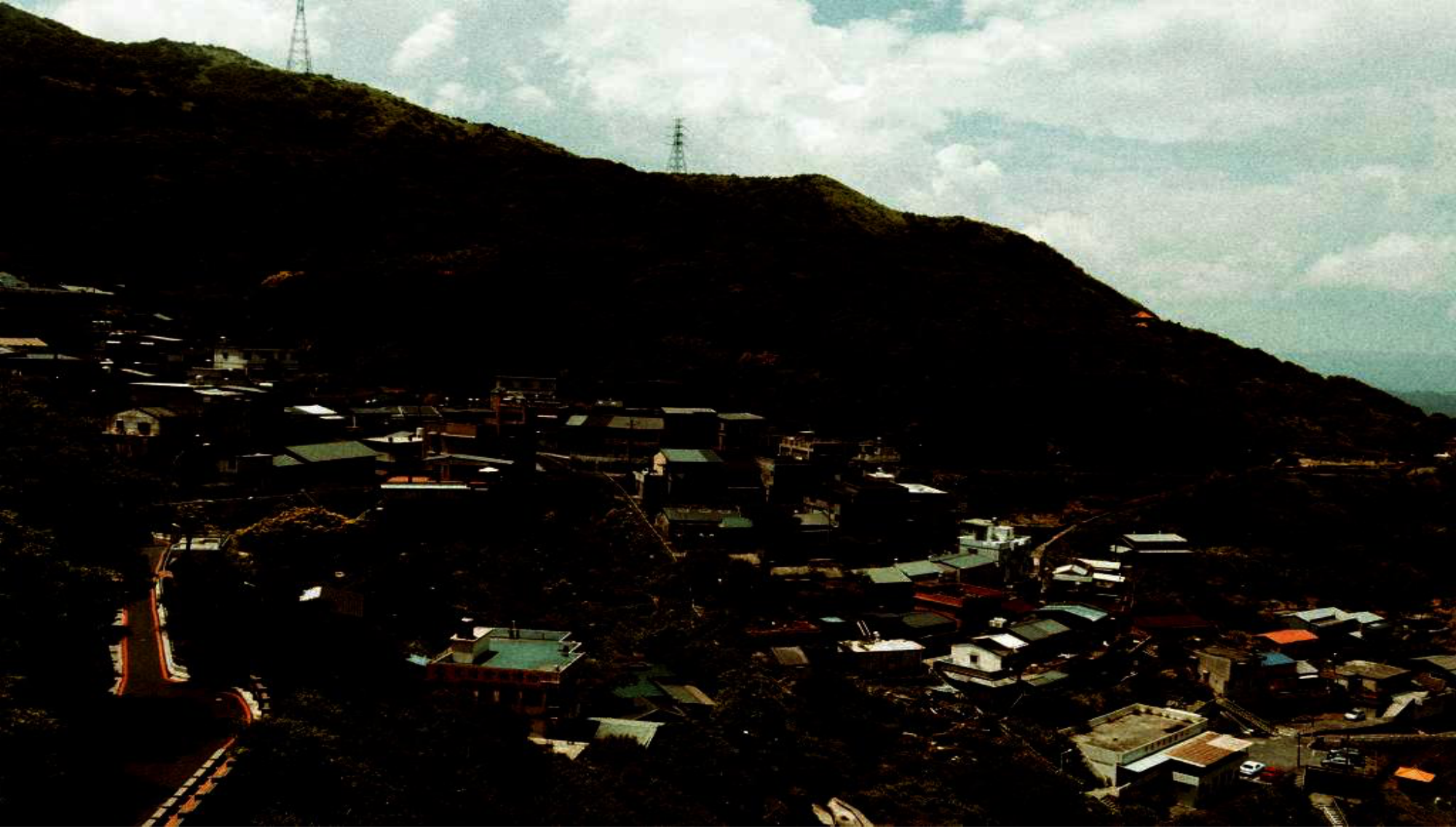}}\hskip.5em 
\end{minipage}
\begin{minipage}[t]{1.0\textwidth}
    \centering
    \vspace{.3em}
    \subfloat[]{\includegraphics[width=0.27\textwidth]{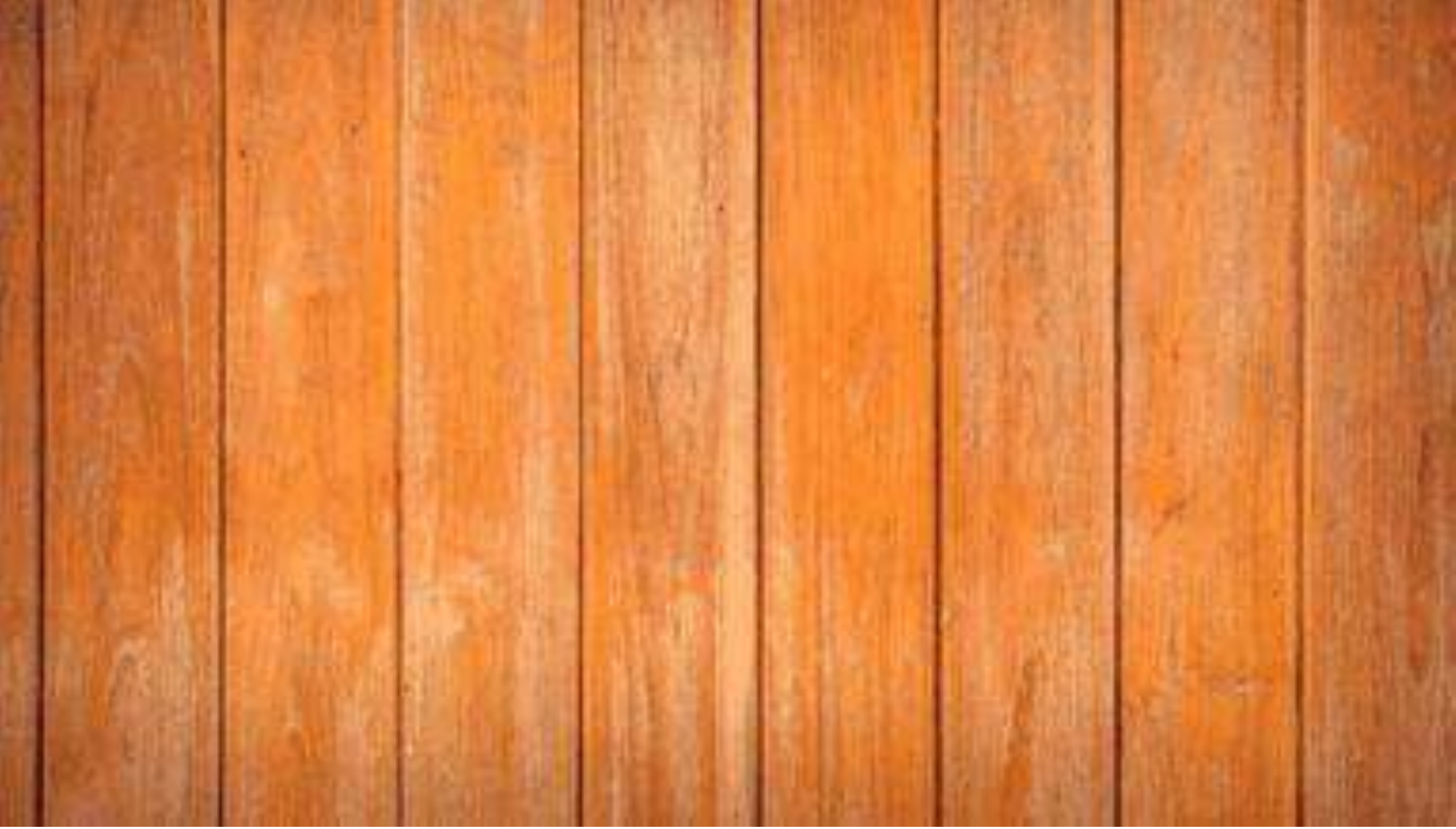}}\hskip.5em
    \subfloat[]{\includegraphics[width=0.27\textwidth]{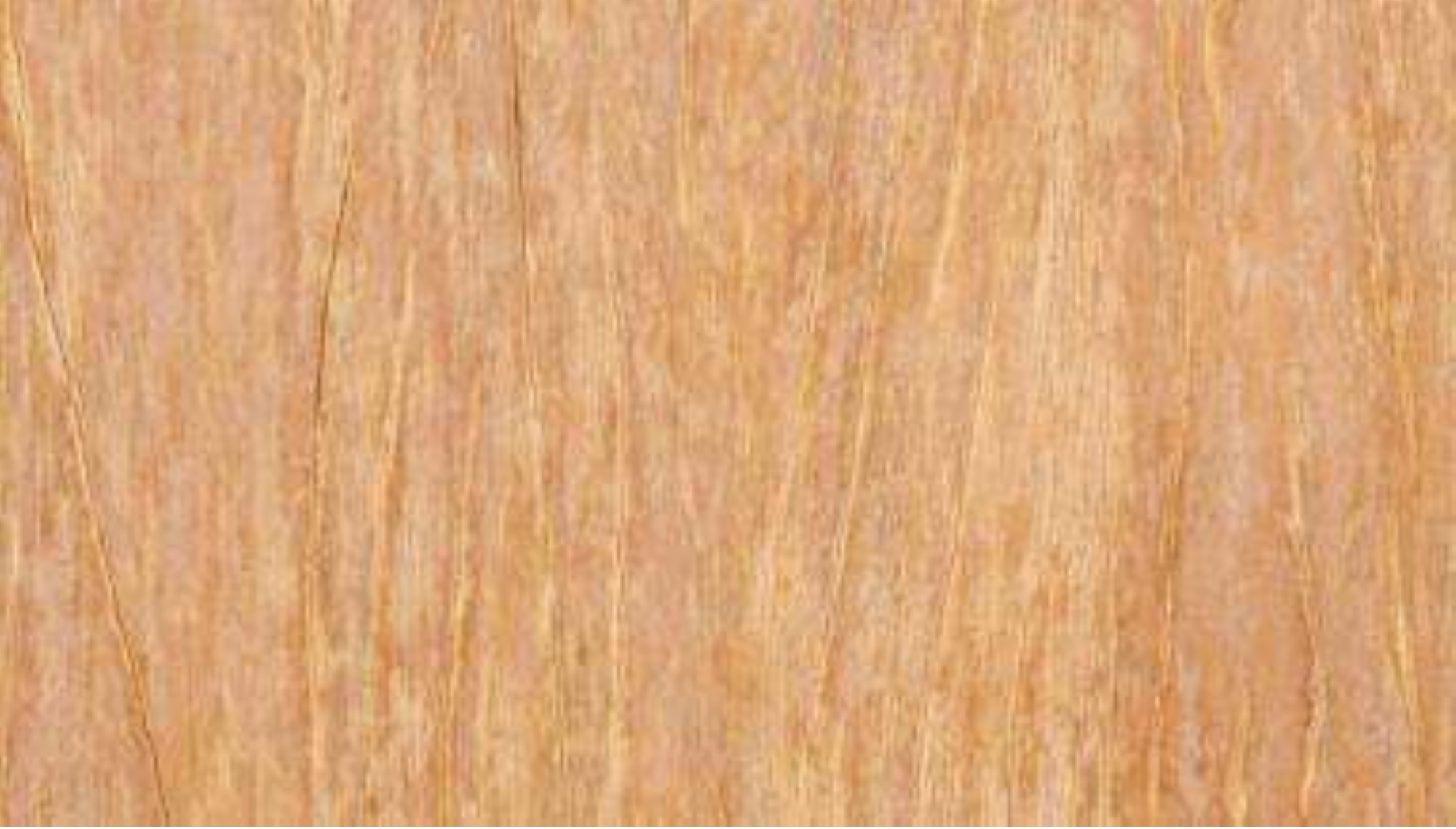}}\hskip.5em
    \subfloat[]{\includegraphics[width=0.27\textwidth]{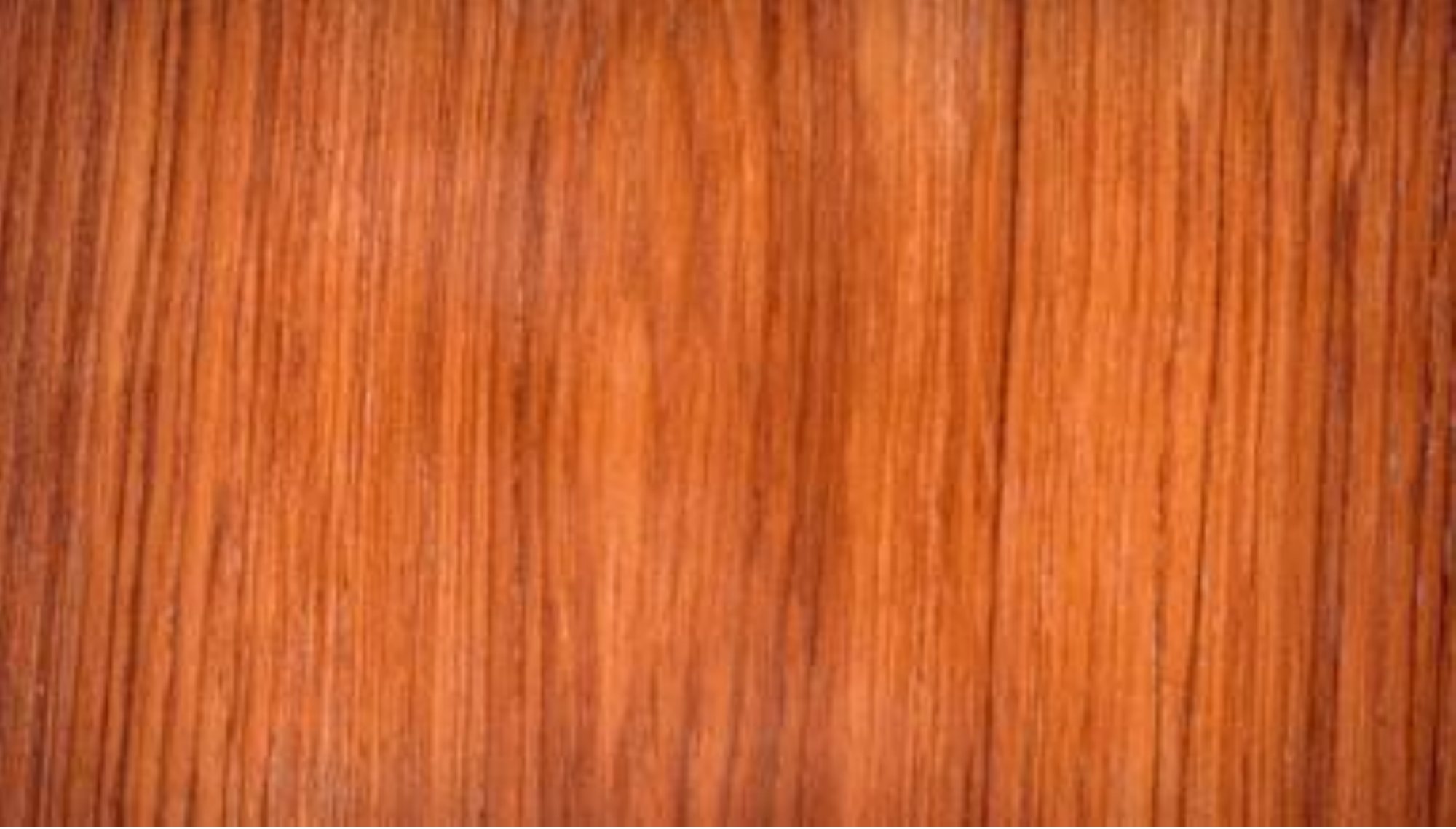}}\hskip.5em
\end{minipage}
\caption{Changes of predictions by our model for images of similar content during the process of active fine-tuning. The predicted value is mapped onto the LIVE MOS scale, with  a higher number indicating better perceptual quality. (a) Image in $\mathcal{L}^{(1)}$ with a predicted value of $76$ before active fine-tuning. (b) Image in $\mathcal{L}^{(2)}$ but not in $\mathcal{L}^{(1)}$ with predicted values of $57$ and $36$ before and after the first round of active fine-tuning, respectively. (c) Image not in $\mathcal{L}^{(1)}$ nor in $\mathcal{L}^{(2)}$  with  predicted values of $69$, $52$, and $52$ before and after the first and the second round of active fine-tuning, respectively. (d) Image in $\mathcal{L}^{(1)}$ with a predicted vaule of $31$ before active fine-tuning. (e) Image in $\mathcal{L}^{(2)}$ but not in $\mathcal{L}^{(1)}$ with predicted values of $44$ and $67$ before and after the first round of active fine-tuning, respectively. (f) Image not in $\mathcal{L}^{(1)}$ nor in $\mathcal{L}^{(2)}$  with  predicted values of $34$, $51$, and $59$ before and after the first and second round of active fine-tuning, respectively.}
\label{fig:model prediction regarding similar images}
\end{figure*}

\subsection{Main Results}
\subsubsection{Quantitative Analysis}
Table~\ref{tab:mainresult} lists the SRCC and PLCC results between model predictions and MOSs on the gMAD image sets $\mathcal{L}^{(1)}$, $\mathcal{L}^{(2)}$, and $\mathcal{L}^{(3)}$, respectively. Before active fine-tuning, all full-reference IQA models surpass the baseline on $\mathcal{L}^{(1)}$, except for SSIM~\cite{wang2004image} in terms of SRCC. After the first round of active fine-tuning on $\mathcal{L}^{(1)}$, our method is able to learn from and combine the best aspects of the competing models, outperforming all of them by a large margin. As expected, the performance of the full-reference models on $\mathcal{L}^{(2)}$ deteriorates. After the second round of active fine-tuning on both $\mathcal{L}^{(1)}$ and $\mathcal{L}^{(2)}$, we do not observe noticeable improvements of our model on $\mathcal{L}^{(3)}$. We speculate that the gMAD examples in $\mathcal{L}^{(2)}$ contain less useful information in refining the proposed method. More importantly, our model may begin to overfit $\mathcal{L}^{(1)}$ and $\mathcal{L}^{(2)}$, as indicated by performance improvements of most full-reference models on $\mathcal{L}^{(3)}$ compared to that on $\mathcal{L}^{(2)}$.  We treat it as a stopping signal of the active fine-tuning cycle. From Table~\ref{tab:mainresult}, it is  interesting to note that the behaviors of the full-reference IQA methods in the gMAD competition are consistent with those on KADID-10k, which shares the same distortion types. When using our method as the anchor in gMAD, we successfully track the progress of full-reference IQA, where the two recent DNN-based models are among the best.

\begin{table}
\caption{Correlation (SRCC and PLCC) of model predictions by $f_w$ against human ratings on $\mathcal{T}$ after simple fine-tuning on $\mathcal{D}_2$  and active fine-tuning on both $\mathcal{D}_2$ and $\mathcal{D}_3$}
\centering
\begin{tabular}{r|cccc}
\toprule
SRCC & LIVE & CSIQ & TID2013 & KADID-10k \\
\hline
Baseline& $0.896$ & $0.859$ & $0.822$ & $0.861$\\
Simple Round 1 & $0.919$ & $0.848$ & $0.826$ & $0.864$\\
Active Round 1 & $0.918$ & $0.863$ & $0.805$ & $0.850$\\
Simple Round 2 & $0.917$ & $0.851$ & $0.835$ & $0.870$\\
Active Round 2 & $0.914$ & $0.871$ & $0.828$ & $0.872$\\
\midrule
\midrule
PLCC & LIVE & CSIQ & TID2013 & KADID-10k \\
\hline
Baseline& $0.915$ & $0.897$ & $0.837$ & $0.866$\\
Simple Round 1  & $0.931$ & $0.891$ & $0.839$ & $0.865$\\
Active Round 1  & $0.930$ & $0.900$ & $0.821$ & $0.858$\\
Simple Round 2  & $0.931$ & $0.896$ & $0.871$ & $0.878$\\
Active Round 2  & $0.931$ & $0.911$ & $0.846$ & $0.881$\\
\bottomrule
\end{tabular}
\label{tab:active_finetuning}
\end{table}

We take a closer look at the performance changes of our method, when it plays the role of the defender and the attacker, respectively.
Fig.~\ref{fig:mean fidelity} shows the mean fidelity losses, where we have several interesting observations. First, after the first round of active fine-tuning, both resistance and aggressiveness  of $f_w$ (in terms of the mean fidelity loss) improve significantly. This suggests that without increasing model capacity (\eg, adding more convolution and GDN layers), our model is able to harness hard gMAD examples. Second, we find that the associated standard errors also reduce, suggesting that the improvements are consistent across a majority of the selected gMAD pairs.  Third, the second round of active fine-tuning slightly improves the resistance, but degrades the aggressiveness of $f_w$, which confirms our previous analysis of potential overfitting. 

Last, we summarize the SRCC and PLCC results of our model on $\mathcal{T}$  in  Table~\ref{tab:active_finetuning}. Noticeable improvements are achieved on all four test sets after two rounds of active fine-tuning. This may be due to two main reasons: 1) more exposure to the training images in $\mathcal{D}_2$ and 2) incorporation of the gMAD image pairs. We conduct an ablation experiment, where we only include images in $\mathcal{D}_2$ for further fine-tuning (see Table \ref{tab:active_finetuning}). 
We find that the first reason is the dominant factor leading to the improvement on $\mathcal{T}$. Therefore, we arrive at a conservative conclusion:
the proposed active learning cycle can be used to improve
the robustness of the BIQA model, without sacrificing the
performance on previously seen data.

\subsubsection{Qualitative Analysis}
We further qualitatively evaluate the   progress of our model in the active fine-tuning cycle. Fig.~\ref{fig:vsi as attacker} shows three gMAD pairs with the maximum fidelity losses (as the worst-case samples) in $\mathcal{L}^{(1)}$, $\mathcal{L}^{(2)}$, and $\mathcal{L}^{(3)}$, respectively, when our model is the defender and VSI~\cite{zhang2014vsi} is the attacker. The pair of images in (a) exhibit dramatically different perceptual quality (in disagreement with our model), 
\textcolor{black}{while those in (b) have closer perceptual quality. This shows that noticeable progress has been made by our model, correcting predictions for strong color distortions.} A  similar result is obtained when MDSI~\cite{nafchi2016mean} attacks our model (see Fig.~\ref{fig:MDSI as attacker}).

We also examine the gMAD image pairs with the maximum fidelity losses, when our model is the attacker. Fig.~\ref{fig:vif as defender} shows the results of VIF~\cite{sheikh2006image} being under attack. The perceptual quality of the images in (a) is close, which is in disagreement with our model. However, the images in (b) are slightly discriminable, indicating that the aggressiveness of our model is improving. Finally, the images in (c) are clearly discriminable, where VIF gives the blurred image less penalty. Fig.~\ref{fig:PieAPP as defender} shows the results of PieAPP~\cite{prashnani2018pieapp} being the defender. Similarly, in $\mathcal{L}^{(3)}$, we successfully identify a strong failure case of PieAPP.

Last, we visualize the changes of predictions by our model on images with similar content, as shown in Fig.~\ref{fig:model prediction regarding similar images}. In the beginning, our baseline model gives high ratings to severely darkened images, while makes low-quality predictions on images of wood textures.  After incorporating images of similar content into the first round of active fine-tuning, our model gives more reasonable predictions to images of similar content not appearing in $\mathcal{L}^{(1)}$. More accurate predictions on images of wood textures can be made after the second round of active fine-tuning. In summary,  we observe a trend that our model adapts  gradually to gMAD examples.


\subsubsection{Further Testing}



\textcolor{black}{In this subsection, we provide further testing of our method. We first  run the D-test, L-test, and P-test \cite{ma2016waterloo} on the Waterloo Exploration Database (with four common distortion types). As shown in Table \ref{tab:ptltest}, we observe performance gains in D-test  and  P-test  after  the  first  round  of  fine-tuning. We believe the improvements would be more signficant if more synthetic distortions are  under test.  After  the  second  round  of  fine-tuning, our model seems to overfit the gMAD examples, as indicated by a slight drop in D-test. Similar phenomenon has also been observed in Table \ref{tab:mainresult}. }

\textcolor{black}{Moreover, we probe the synthetic-to-real generalization of our method  on two authentically distorted  datasets - SPAQ \cite{fang2020cvpr} and KonIQ-10k \cite{hosu2020koniq}. Table \ref{tab:spaq_and_koniq} shows the results in term of SRCC and PLCC. We find that active fine-tuning from gMAD examples improves the synthetic-to-real generalization of BIQA models, which provides additional justification of the proposed method.}


\begin{table}[t]
  \centering
  \caption{Results of the D-test, L-test, and P-test on the Waterloo  Exploration  Database}
  \label{tab:ptltest}
  \begin{tabular}{l|cccc}
     \toprule
     & D-test  & L-test & P-test\\
     \hline
     Baseline & $0.881$ & $0.981$ & $0.998$ \\
     Round1  & $0.890$ & $0.969$ & $0.999$\\
     Round2 & $0.881$ & $0.975$ & $0.999$\\
     \bottomrule
  \end{tabular}
\end{table}

\section{Conclusion and Discussion}
We have introduced an active fine-tuning cycle for improving BIQA methods. Combining with the training techniques for constructing the baseline, we have presented a complete and practical framework to learn a top-performing BIQA model that 1) relies on only a handful of human-labeled images, 2) delivers superior performance on existing IQA databases of synthetic distortions, and 3) exhibits strong aggressiveness and resistance in gMAD, even when competing with a set of full-reference IQA methods. 

We used the gMAD competition to seek informative samples for active fine-tuning. It is of interest to examine whether traditional query strategies~\cite{settles2009active}, such as those based on uncertainty sampling, expected model change and expected error reduction, can facilitate the robustness of the BIQA model, and to compare the results with ours under the same human-labeling budget. Recently, Wang~\etal~\cite{wang2020iamgoingmad} extended the idea of gMAD to compare a number of ImageNet classifiers. It is thus natural to explore the current work in the context of image classification  as a way of improving the generalization of the classifiers to natural image manifold.

Our work presents a new line of research in BIQA. We conclude by listing other research directions that, we believe, are worth exploring. First, it is desirable to adapt BIQA models trained on a fixed set of synthetic distortion types to unseen ones. Xu~\etal~\cite{xu2016blind} made one of first attempts by exploiting higher order image statistics. Second, a practical BIQA model should be able to handle both synthetic and realistic camera distortions. It is interesting to extend our work to such a cross-distortion-scenario setting. Third, a universal BIQA method should embody a prior  probability model of natural undistorted images. Mittal~\etal~\cite{mittal2013making} developed such a model with reasonable generalizability. Fourth, how to incorporate high-level semantics into the design of BIQA is yet another challenging problem for future research.

\begin{table}[t]
  \centering
  \caption{Synthetic-to-real generalization of our method on two authentically distorted  datasets - SPAQ \cite{fang2020cvpr} and KonIQ-10k \cite{hosu2020koniq} 
  }\label{tab:spaq_and_koniq}
  \begin{tabular}{l|cc}
     \toprule
     SRCC & SPAQ  & KonIQ-10k  \\
     \hline
     Baseline & $0.575$ & $0.403$\\
     Round 1 & $0.600$ & $0.523$\\
     Round 2 & $0.633$ & $0.533$\\
     \hline
     \hline
     PLCC & SPAQ & KonIQ-10k\\
     \hline
     Baseline & $0.579$ & $0.426$ \\
     Round 1 & $0.609$ & $0.526$ \\
     Round 2 & $0.635$ & $0.532$ \\
     \bottomrule
  \end{tabular}
\end{table}


\ifCLASSOPTIONcaptionsoff
  \newpage
\fi

\bibliographystyle{IEEEtran}
\bibliography{main}


\end{document}